\documentclass[12pt]{article}
\pdfoutput=1
\usepackage{setspace}
\usepackage{color}
\usepackage{amsmath, amssymb,slashed,url}
\usepackage{amsthm}
\usepackage{graphicx,epsfig}

\numberwithin{equation}{section}

\def\a{\alpha}
\def\b{\beta}
\def\r{\gamma}
\def\d{\delta}

\def\m{\mu}

\def\p{\pi}
\def\lam{\lambda}
\def\w{\omega}

\def\S{\Sigma}
\def\slr{SL(2,\mathbb{R})}
\def\slc{SL(2,\mathbb{C})}
\def\Tr{\mbox{Tr}\,}
\def\pd{\partial}
\def\td{\tilde}
\def\tch{Teichm\"uller}
\def\beq{\begin{equation}}
\newcommand{\eeq}[1]{\label{#1}\end{equation}}
\def\bea{\begin{eqnarray}}
\newcommand{\eea}[1]{\label{#1}\end{eqnarray}}
\def\ba{\begin{array}}
\def\ea{\end{array}}
\renewcommand{\Im}{\mbox{Im}\,}
\renewcommand{\Re}{\mbox{Re}\,}
\date{}
\begin{document}
\def\draftnote#1{{\color{red}#1}}
\begin{titlepage}
\hfill CERN-PH-TH-2015-188 
\vspace{20pt}
\begin{center}
{\huge   On a Canonical Quantization of 3D Anti de Sitter Pure Gravity}

\vspace{18pt}

{\large  Jihun Kim$^a$  and Massimo Porrati$^{a,b}$}

\vspace{12pt}

{$^a$ \em Center for Cosmology and Particle Physics, \\ Department of Physics, New York University, \\4 Washington Place, New York, NY 10003, USA}

\vspace{12pt}

{$^b$ \em CERN PH-TH,
CH 1211, Geneva 23, Switzerland\footnote{Until September 1, 2015; on sabbatical leave from NYU.}}

\end{center}

\abstract{
We perform a canonical quantization of pure gravity on $AdS_3$ using as a technical tool its equivalence at the 
classical level with a 
Chern-Simons theory with gauge group $\slr\times\slr$. We first quantize the theory canonically on an asymptotically AdS
space --which is topologically the real line times a Riemann surface with one connected boundary. Using the 
``constrain first'' approach we reduce canonical quantization to quantization of orbits of the Virasoro
group and K\"ahler quantization of Teichm\"uller space. After explicitly computing the K\"ahler form for the torus with one boundary
component and after extending that result to higher genus, we recover known results, such as that wave functions of 
$\slr$ Chern-Simons theory are 
conformal blocks. We find new restrictions on the Hilbert space of pure gravity by imposing invariance under 
large diffeomorphisms and normalizability of the wave function. The Hilbert space of pure gravity is shown to be the target
space of Conformal Field Theories with continuous spectrum and a lower bound on operator dimensions. 
A projection defined by topology changing amplitudes in Euclidean gravity is proposed. It defines an invariant subspace 
that allows for a dual
interpretation in terms of a Liouville CFT. Problems and features of the CFT dual are assessed and a new definition of the
Hilbert space, exempt from those problems, is proposed in the case of highly-curved $AdS_3$.}

\newpage
\pagestyle{empty}
\tableofcontents
\begin{quotation}
{\it This whole book is but a draught--nay, but the draught of a draught. Oh, Time, Strength, Cash, and Patience!}\end{quotation}
\end{titlepage}
\section{Introduction}\label{intro}
Canonical quantization of gravity has a long history that even pre-dates 
the classic 1967 DeWitt paper~\cite{wdw} (e.g.~\cite{adm}). Since 
then, books~\cite{books}, historical reviews~\cite{rov}, learned and lively debates~\cite{in-out} have been published on the
 subject. Its status is still mired in some of the very same problems of half a century ago --one of which, somewhat ironically, is the
absence of an unambiguously defined time evolution in closed universes. 
It may seem hard to have anything new to add to the subject. For pure gravity in 3D it may seem even hopeless. After all,
this is a case where no local degrees of freedom propagate and the Einstein action --with or without cosmological constant-- can be re-written as a Chern-Simons (CS) action~\cite{at,w}. Yet, pure gravity in 3D Anti de Sitter (AdS) space is 
nontrivial because it allows for (BTZ) black hole solutions~\cite{BTZ} and boundary gravitons.... and yet we will be able 
to say a few new things 
about the canonical quantization of precisely such a theory. 

Before discussing any new result we must recall a few old ones, as well as the definition of the theory itself. 

\subsection{The Theory}

3D pure gravity propagates no bulk gravitons so it is in principle possible to quantize it by first solving the non-dynamical 
equations that generate constraints on the space of solutions --i.e. the Gauss laws of general covariance-- 
and subsequently quantize the space of physical states. This ``constrain first procedure'' is not guaranteed to be 
equivalent to the ``quantize first'' one in general. In the case of pure 3D gravity 
an additional well known property may help understanding the quantization procedure in either approach: pure gravity is classically equivalent to 
a Chern-Simons theory~\cite{at,w}. When the cosmological constant is negative, the appropriate Chern-Simons theory is based on the gauge group $SL(2,\mathbb{R})\times SL(2,\mathbb{R})$. Here we are ignoring subtleties related to global 
properties of the gauge group so that we do not distinguish between, say, $SL(2,\mathbb{R})\times SL(2,\mathbb{R})$ and 
$SO(1,2)\times SO(1,2)$. Some of these questions will be addressed when we will examine the role of global differmorphisms.

In the ``constrain first'' approach, the space of classical solutions
of either the second-order Einstein gravity or the first-order Chern-Simons theory coincide and thus give rise to
equivalent quantizations. While not obvious, we will present an argument supporting such equivalence. Moreover, for some Chern-Simons theories, specifically for the $SL(2,\mathbb{R})$ 
theory, it was shown that the ``quantize first'' approach gives rise to the same vector space as the ``constrain first" approach. 

The previous statements must be qualified. As we will see explicitly later, on a product manifold of the form 
$\Sigma\times \mathbb{R}$,  with $\Sigma$ a Riemann surface, the Chern-Simons Gauss law is solved by flat
connections on $\Sigma$. This space is much larger than the space of acceptable Einstein metrics on 
$\Sigma\times \mathbb{R}$. In other words, the moduli space of the Chern-Simons formulation of 3D gravity is too large. 
Moreover, the ``good" part of the moduli space and the bad one can be connected by a path passing 
through only mildly singular geometries. After quantization, a wave function could
``spill" through the singularity from one sector to the other. Forbidding such pathology will give us the most important constraint on the 
quantized theory.

The Chern-Simons formulation is very useful for studying pure  three-dimensional gravity because it simplifies the 
non-dynamical constraint equations. So, in the rest of this section, we briefly review three dimensional gravity in the Chern-Simons formulation, we summarize general
properties of K\"ahler quantization, we review the ``quantize first'' approach used in~\cite{V,VV} for quantizing
$SL(2,\mathbb{R})$ Chern-Simons theory and we describe some features of the ``constrain first" approach. 

We will also review relevant aspect of quantization of Teichm\"uller theory. The next two subsections present background materials for 
completeness and to fix notations. Subsections 1.4 and 1.5 will summarize the new features that distinguish AdS$_3$ 
quantum gravity from a theory of quantum \tch\ space. 

\subsection{Chern-Simons Formulation}
The Einstein-Hilbert action in the first order formalism  is given by

\begin{equation}
I_{EH} = \frac{1}{16 \pi G} \int{ \left[ e^a \wedge \left( 2 d  \omega_a - \epsilon_{abc} \omega^b \wedge \omega^c \right) + \frac{1}{3 l^2} \epsilon_{abc} e^a \wedge e^b \wedge e^c  \right]  }
\end{equation}
where $\omega_a = \frac{1}{2} \epsilon_{abc} \omega^{bc}$ and $ -1/l^2 \equiv \Lambda $ is the cosmological constant .
Up to a boundary term, this action can be decomposed into two copies of an $SL(2,\mathbb{R})$ Chern-Simons action:

\beq
I_{CS} = \frac{k}{4 \pi } \int{ \mathrm{Tr}  \left( A^+ \wedge dA^+ + \frac{2}{3}  A^+ \wedge A^+ \wedge A^+ \right) } - \frac{k}{4 \pi } \int{ \mathrm{Tr}  \left( A^- \wedge dA^- + \frac{2}{3} A^- \wedge A^- \wedge A^- \right)}.
\eeq{cs}
In equation~(\ref{cs}), the gauge fields are defined as $A^{\pm}_a = \omega_a \pm \frac{1}{l} e_a$, and the coupling constant $k$ is related to $G$ by $k = \frac{l}{4 G}$. The addition of a gravitation CS term makes the first order
theory different from the second order one. In the former case, generically gravity has extra (ghost) 
excitations~\cite{deser} while in the latter the classical theory is the same as pure gravity and is described by two $\slr$ CS theories with different level $k\neq k'$.  For this reason we are confining ourselves to the $k=k'$ case from now on. 

The equations of motion following from $I_{CS}$ state that the gauge connections are flat
\begin{equation}
F^{\pm} = dA^{\pm} + A^{\pm} \wedge A^{\pm} = 0.
\end{equation}
When the dreibein $e^a$ is invertible, this is equivalent to the constant curvature and torsion free conditions
\begin{eqnarray}
d\omega + \omega \wedge \omega &=& -{1\over l^2} e\wedge e 
\\
de + \omega \wedge e &=& 0.
\end{eqnarray}
 Infinitesimal general coordinate transformations $\epsilon^\mu$ act as gauge transformations on $A^\pm$ precisely when the gauge  connection is flat, because of the equation
 \beq
 \delta_{gct}A_\mu= \epsilon^\nu F_{\nu\mu} + D_\mu (\epsilon^\nu A_\nu) = D_\mu (\epsilon^\nu A_\nu) \mbox{ at } 
 F_{\mu\nu}=0.
 \eeq{gct}
 So, two solutions of Einstein's equations, which are  equivalent under small diffeomorphisms, are also gauge equivalent when 
 seen as $SL(2,\mathbb{R})\times SL(2,\mathbb{R})$ flat connections. When the 3D local frame $e^a$ is invertible the converse 
 also holds. Notice that eq.~(\ref{gct}) does not guarantee that the equivalence extends to large diffeomorphisms. 
 
Concluding that Einstein's action of three dimensional pure gravity is classically equivalent to two copies of the 
Chern-Simons action would be nevertheless incorrect. After all, there exist many solutions of Chern-Simons theory 
that cannot
be interpreted metrically; the most obvious example being $A^{\pm}=0$ everywhere. The best we can hope for is that a well defined subspace of solutions of Chern-Simons theory be equivalent to the space of solutions of pure gravity. 
When $\Lambda < 0$ there are strong indications, which we will review later in the paper, that this is indeed the case. 
So, for the
time being, we will work with the Chern-Simons action. Since the action is a sum over two independent $SL(2,\mathbb{R})$ gauge connections, we start by studying one of the copies.

To quantize the problem canonically, we must assume that the original three-manifold $M$ is  
$\S \times \mathbb{R}$, where $\Sigma$ is a Riemann surface.
In this approach, by denoting differential form and gauge field over the Riemann surface as $\td{d}$ and $\td{A}$, the action can be written as
\begin{equation}\label{action_A}
I_{CS} = -\frac{k}{4 \p} \int_\mathbb{R}{dt \int_{\S}{ \mathrm{Tr} \left( \td{A} \pd_t\td{A} \right)}} + \frac{k}{2 \p} \int_\mathbb{R}{dt\int_{\S}{\mathrm{Tr} \left[ A_0 \left(  \td{d}\td{A} + \td{A} \wedge \td{A} \right) \right] }},
\end{equation}
up to a boundary term proportional to $A_0$. This term is canceled by choosing $A_0|_{\partial\Sigma}=0$.
The field $A_0$ is a Lagrange multiplier rather than a dynamical variable: by varying action (\ref{action_A}) with respect to $A_0$, we get the constraint equations
\begin{equation}\label{const}
\td{d} \td{A} + \td{A} \wedge \td{A} = 0.
\end{equation}
So, $\tilde{A}$ is a flat connection on $\Sigma$ and the classical phase space, ${\cal M}$, is the space of flat connections modulo
gauge transformations. For compact gauge groups, most of the properties of quantum Chern-Simons theory were 
found in~\cite{Wit}. For our purposes, we will need two modifications of the compact-group case: first of all, our group 
is noncompact. Second, our space has a boundary. 

To take into account the presence of the boundary, let us consider the solution of the constraint~(\ref{const}). Its general solution is given by
\begin{equation}
\tilde{A} = g^{-1} \td{d} g + g^{-1} H g ,
\end{equation}
where $g$ is a single valued function from the Riemann surface $\S$ to the group $G$, and $H$ is a yet to be found family of connections
specified by holonomies around the non-contractible cycles of $\S$.
The flat-connection condition is $\td{d} H + H \wedge H = 0$.
With the help of this, we can reduce action (\ref{action_A}) on flat connections to
\begin{eqnarray}
I_{CS} &=& \frac{k}{4 \p}\int_{\mathbb{R}}{dt \int_{\pd \Sigma}{ \mathrm{Tr} \left( g^{-1} \td{d} g g^{-1} \dot{g} \right) }} + \frac{k}{12 \p}\int_{M}{\mathrm{Tr} \left( g^{-1} {d} g \right)^3} \nonumber
\\
&\quad& -\frac{k}{2 \p}\int_{\mathbb{R}}{dt\int_{\pd \Sigma}{\mathrm{Tr}\left( g^{-1} \dot{g} H \right)}} - 
\frac{k}{4 \p} \int_{\mathbb{R}}{dt\int_{\S}{\mathrm{Tr} \left( H \wedge \dot{H} \right)}} .
\label{action_eff}
\end{eqnarray}
The first line of this equation is the action for a chiral WZW model. Its canonical quantization was studied in~\cite{MS}. We will review it in more details later on. The first term in the second line of~(\ref{action_eff}) introduces one extra degree
of freedom in the theory, since the reparametrization $g\rightarrow g \omega(t)$ is a gauge invariance for the WZW action
but not for action~(\ref{action_eff}). As shown in~\cite{MS}, this extra degree of freedom defines which representation of the
loop group $\hat{G}$ appears in the quantum Hilbert space of action~(\ref{action_eff}). Again, this statement and the next one
will be explained in details later. The last term in~(\ref{action_eff}) defines a symplectic structure on a
 finite-dimensional phase space. 
 One component of such space  is the Teichm\"uller space of $\S$. 
 Its treatment is the subtlest part of the quantization program. In particular, certain constraints arising from normalizability 
 of the $\slr\times\slr$ wave function and  invariance of quantum gravity under large 3D diffeomorphisms will gives rise to new and 
 nontrivial  properties of the quantum theory.  
 
\begin{figure}[h]
\begin{center}
\epsfig{file=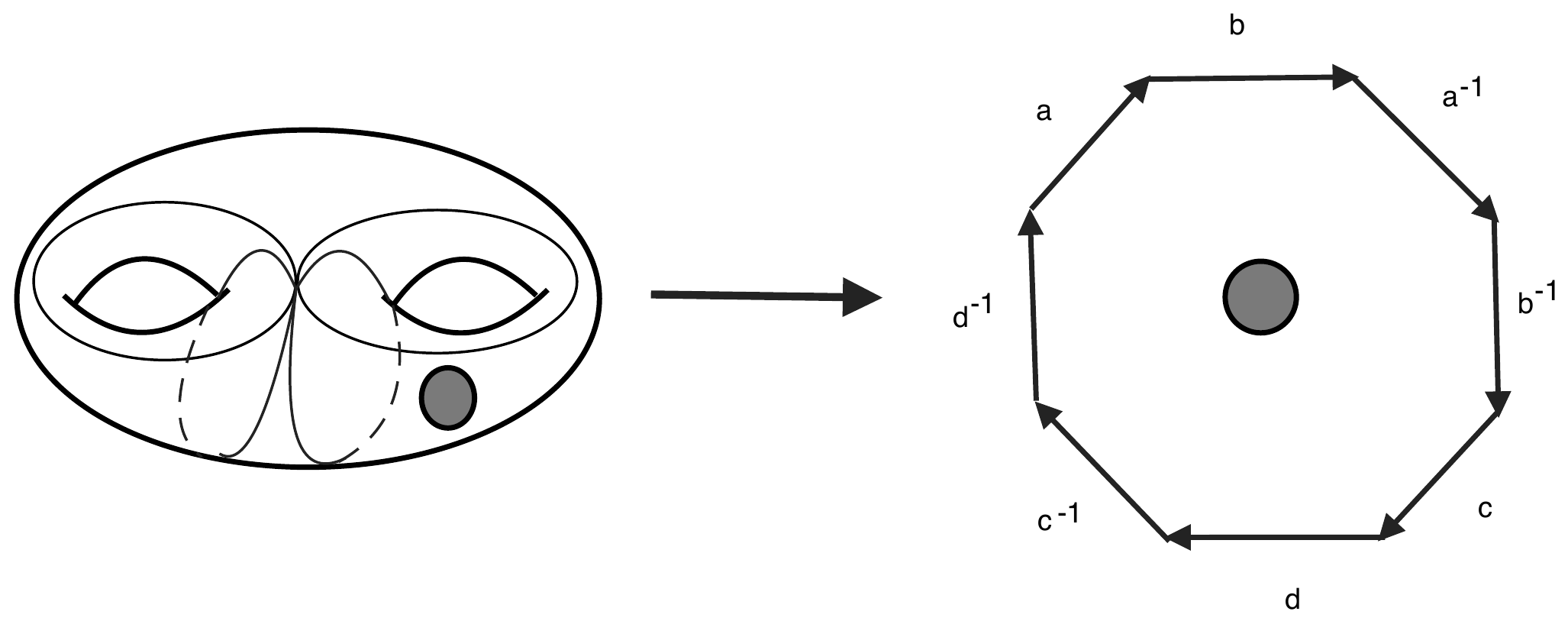, height=2in, width=5in}
\end{center}
\caption{Cutting a surface $\S$ with a single boundary along homotopy cycles produces a disk with a hole (the grey disk).}
\label{cutsurface}
\end{figure}

 To compute the symplectic structure we will find it useful to cut the surface along the closed path shown  in 
 figure~(\ref{cutsurface}). It is made of two components: the first is a canonical set of homotopy cycles, the second 
is the boundary, i.e. the edge of the hole.  Cutting the surface along such path 
 we define in our case a disk with a hole (see e.g.~\cite{far-kra}).  
  
\subsection{Quantization of Teichm\"uller space}

Quantization of $SL(2,\mathbb{R})$  Chern-Simons theory on a punctured Riemann surface $\S$ was studied first in~\cite{V} (see also~\cite{VV}) in the ``quantize
first" approach.  Refs.~\cite{V,VV} argued that physical wave functions obeying the Gauss law constraints  of 
$SL( 2, \mathbb{R} )$ Chern-Simons theory are Virasoro conformal blocks~\cite{cb} and provide a quantization of  the Teichm\"uller 
space of the surface $\S$. This conjecture was proven for genus zero surfaces in~\cite{tesch} and for all 
genera in~\cite{tesch2}. See also~\cite{mod} for related work and~\cite{flat} for a recent review. 
It plays an essential role in generalizations of the
AGT~\cite{agt} correspondence~\cite{agt1,agt2}.  In fact, a similar identification holds for $SL(n,\mathbb{R})$ and $SL(n,\mathbb{C})$ Chern-Simons theories and higher moduli spaces~\cite{hitch-q}. A related quantization procedure for
flat $\slc$ connections was discussed in~\cite{shat}. 

Let us review first the quantization of Teichm\"uller space following the original approach of~\cite{V} and \cite{VV}.
The  gauge potentials are written in an $SU(1,1)$ basis as $A=i\sigma_3 \w + e^+ \sigma^- +
e^-\sigma^+$, with $\w,e^\pm$ real. Since the curvature is defined as $F=dA +A\wedge A$, $\w$ is the gauge field 
associated to the compact generator of $SU(1,1)$. This decomposition defines a 2D local frame (the ``zweibein'') 
$\left( e^+ , e^- \right)$ and an $SO(2)$ spin connection $\w$. The Gauss law constraint equations are
\beq
\mathcal{G}^{\pm} = de^{\pm} \mp \w \wedge e^{\pm}  = 0 , \qquad
\mathcal{G}^0 = d\w + e^+ \wedge e^- =0.
\eeq{constr}

On the solution of constraints~(\ref{constr}), when the local frame $e^\pm$ is invertible, $SU(1,1)$ 
gauge transformations are equivalent to local Lorentz transformations, $LL$, and diffeomorphisms connected to the identity, $Diff_0$. The constraint equations imply that the 2D metric with connection $\w$ and metric $ds^2=e^+ \otimes e^-$ has constant curvature.
The space of flat connections is then given by  $ \{e^\pm,\w | d\w(e)=-e^+ \wedge e^-\}/(LL \times Diff_0)$, where $\w(e)$ solves the constraints $G^\pm=0$. Since the set of all
constant-curvature local frames is quotiented by $LL\times Diff_0$, but not by large diffeomorphisms, 
the coset is Teichm\"uller space rather than the moduli space of $\S$. 

Quantization means that classical variables are promoted to non-commuting coordinates and momenta. In the case of Chern-Simons theory with a compact gauge group, there is a natural procedure for quantizing the theory ``upstairs,'' that is before imposing the Gauss law constraints~\cite{Wit,dpw}. In the compact case, a complex structure on $\S$ induces canonically a complex structure on the gauge fields, which are thus divided into holomorphic coordinates $A_z$ and antiholomorphic coordinates $A_{\bar{z}}$.
Such structure is in fact a positive K\"ahler structure with K\"ahler potential 
$K= -\int d^2 z \,\Tr A_z A_{\bar{z}}$ ($A$ is antihermitian). The K\"ahler metric is invariant under gauge transformations and it  induces a K\"ahler
structure on ${\cal M}$, the moduli space of flat connections.  
On a K\"ahler space, quantization is done in analogy with coherent state quantization for the harmonic oscillator: 
the K\"ahler structure on a $n$-manifold $M_n$, covered by an atlas $\{U_i\}$, defines locally in every 
chart K\"ahler potentials $K_i$.
On the intersection of two charts $K_i|_{U_i\cap U_j}=K_j|_{U_i\cap U_j}+ F_{ij} +\bar{F}_{ij}$ with $F_{ij}$ holomorphic. 
Wave functions are holomorphic sections of the line bundle defined by the transition function $F_{ij}$; normalizable in the norm
\beq
|| \psi ||^2= \int_{M_n} \psi(z)\wedge * \overline{\psi(z) }\exp(-K).
\eeq{m1aa}

For non-compact groups, such procedure fails because the K\"ahler
metric is either gauge invariant or positive, but not both. For $SU(1,1)$, ref.~\cite{V} defined an 
$SU(1,1)$-invariant K\"ahler structure which, while non positive ``upstairs," becomes positive on 
the (Teichm\"uller component of) the moduli space of flat connections. 

To proceed with the ``quantize first'' approach of refs.~\cite{V,VV}, it is convenient to parametrize the local frame as~\cite{V}
\beq
e^+ = e^{\varphi} \left( d z + \m  d \bar{z} \right) , \quad e^- = e^{\bar{\varphi}} \left(\bar{\m} d z + d  \bar{z} \right).
\eeq{m1a}
One must define next Poisson brackets.
Since the Chern-Simons action is first-derivative in time, the gauge potentials represent both coordinate and conjugate momenta.
In the case at hand, ref.~\cite{V} interpreted $e^+_z$, $e^+_{\bar{z}}$ and $\w_z$ as coordinates and $e^-_{\bar{z}}$, $e^-_z$ and $\w_{\bar{z}}$ as conjugate momenta. The non-vanishing Poisson brackets are
\bea
\{ e^+_z \left(z\right) , e^-_{\bar{w}}\left(w\right) \} &=&  \frac{4\p}{k} \d \left(z-w\right),  \nonumber \\
\{ e^+_{\bar{z}}\left(z\right) , e^-_z \left(w\right) \} &=& -\frac{4\p}{k} \d \left(z-w\right), \nonumber \\
\{\w_z \left(z\right) , \w_{\bar{w}}\left(w\right) \}  &=& -\frac{4\p}{k} \d \left(z-w\right).
\eea{poissbra}

Quantizing the theory in the ``quantize first'' approach means promoting the Poisson brackets to quantum commutation relations and promoting the classical constraint equations to operators that must annihilate physical states.
The constraints ${\cal G}^+$ and ${\cal G}^0$ can be solved exactly, resulting in the following form for a physical wave function
\bea
\Psi \left[ \varphi  , \m , \w \right] &=& e^{W\left[ \w, \varphi, \m\right]} \psi \left[ \m \right], \nonumber \\
W\left[ \w , \varphi, \m \right] &=& \frac{k}{4\p} \int{\left[ \frac{1}{2} \pd \varphi \bar{\pd} \varphi - \w \pd \varphi - \m \left( \frac{1}{2} \left( \pd \varphi - \w \right)^2 - \pd \left( \pd \varphi - \w \right) \right) \right]}.
\eea{phys-wf}
The action of ${\cal G}^-$ on $ \psi \left[ \m \right] $ then reduces to the Virasoro Ward identity~\cite{V}.

To find the inner product between physical states one follows the general rule of K\"ahler  quantization. So the inner 
product $ \langle  \Psi \vert  \Psi \rangle$ is determined by the K\"ahler potential, which in our case is  
\beq
K= \frac{k}{4\p}\int{\left(  \w_z \w_{\bar z}  - e^+_z e^-_{\bar z} \right) }, \qquad e^-_z=\overline{e^+_{\bar{z}}}.
\eeq{K-ver}

 Notice that $K$ is invariant under $SU(1,1)\sim SL(2,\mathbb{R})$, but it is not positive definite. On the physical wave 
 functions~(\ref{phys-wf}), on the other hand, the scalar product is positive definite and it is given by~\cite{V,VV}
\begin{eqnarray}
\langle \Psi_1 \vert \Psi_2 \rangle_V   &=&  \int{ \left[d\w d\varphi d\m\right] e^{ -K + \overline{W}_1 + W_2 } \bar{\psi}_1\left[\bar{\m}\right] \psi_2 \left[\m\right] } \nonumber \\
&=& \int{ \left[ d\varphi d\m\right] e^{S_L\left[\phi,\m,\bar{\m}\right] - {\cal K}\left[\m,\bar{\m}\right] }}\bar{\psi}_1\left[\bar{\m}\right] \psi_2 \left[\m\right] .\label{inner_ver}
\end{eqnarray}
In equation (\ref{inner_ver}), $S_L$ is the Liouville action with central charge $26-c$ (notice that the Liouville theory is time-like for large $c$), where $c=6k$, and ${\cal K}$ is the so-called holomorphic anomaly term~\cite{V}. This term can be understood in a different way as part of another 
Liouville action, the latter defined not only by data on the Riemann surface $\S$ but also by a 3D surface $M$ such that
$\Sigma=\partial M$~\cite{zo-tak,tak-teo,tak-lap}. We will find that the K\"ahler potential  in our ``constrain first'' approach 
is also given essentially by $-S_L+{\cal K}$; different definitions of the 
Liouville action will be discussed in some detail later on. 

The quantization sketched above can be done in much greater rigor by using the ``constrain first" 
approach, which reduces  the moduli space of flat $SL(2,\mathbb{R})$ connections to a 
finite-dimensional (Teichm\"uller) space endowed with a K\"ahler metric. This procedure also makes it clear how to 
implement the metric non-degeneracy condition $\det e <0$.

Quantization of Teichm\"uller spaces of compact and punctured surfaces has progressed enormously in recent times, thanks to new 
coordinate systems~\cite{fock,che-fo,kash} particularly useful for quantization. In most of the modern literature on the
subject, an algebraic approach is followed: the quantization problem is reduced to studying an algebra of operators subject
to polynomial constraints and its representation. 
We found it convenient to use instead the different though related~\cite{tesch,tesch2}  approach of directly
defining the Hilbert space by finding a norm in the  ``coherent state'' basis defined by the K\"ahler structure. This 
approach is useful for us since we are less concerned here with explicit realizations of the operator algebra 
than with finding generic links 
between the quantization of 3D gravity and 2D conformal field theories. We want in particular to see if
holography may hold for canonically quantized pure gravity. 

Another difference with the $SL(2,\mathbb{R})$ case usually discussed in the literature 
is that we deal with a Riemann surface with a boundary.
The Teichm\"uller space is then infinite dimensional, but, as we mentioned already
 in the previous subsection, the infinite-dimensional part
of the space is in some sense trivial to quantize, while the nontrivial part is finite-dimensional. 

The most important difference though is that 3D gravity must be invariant also under large diffeomorphisms. Therefore, 
we shall (and will)  study in details the new constraints that follow from invariance under global diffeomorphisms. 
Even for understanding these constraint, K\"ahler quantization will turn out to be a most useful tool. Here it
will suffice to mention the upshot of our analysis: the phase space of pure AdS$_3$ gravity is $[T(\S)\times T(\S)] / M$, 
that is two copies of Teichm\"uller space quotiented by {\em a single} copy of the mapping class group $M$ 
of the Riemann surface $\S$, which acts diagonally. 
We must note at this point that, modulo the presence of a boundary, this is the space
predicted by Witten to arise from canonical quantization of AdS$_3$ Chern-Simons gravity~\cite{w8}. 

\subsection{More on Quantization}

Wave functions on the coset $[T(\S)\times T(\S)] / M$ must be invariant under $M$. This means that the factorized vector
$\psi(\mu)\phi(\bar{\mu}')$, $\mu\in T, \mu'\in T'$ is not a physical state.\footnote{The complex conjugation in
the definition of the second wave function simplifies its comparison with CFT objects that will be studied in section 5.}  
The correct $M$-invariant state is in general non factorizable. To find its properties we must recall how $\psi(\mu)$ transforms under $M$. Generally, the K\"ahler potential is
not invariant but instead transforms under an element of $M$ as 
\beq
K\rightarrow K'= K + F + \bar{F},
\eeq{k-t} 
with $F$ a holomorphic function of $\mu$. To ensure invariance of the scalar product, a wave function $\psi_I$ must 
thus transform as $\psi_I \rightarrow \psi'_I= \exp(F)U_I^J \psi_J$. The indices $I,J$, which can be either discrete or 
continuous, are raised and lowered with a positive metric and the matrices $U$ are unitary: 
$\sum_LU_I^L \overline{U_J^L}=\delta_I^J$. Explicit forms of $U$ are known~\cite{tesch,mod} but the K\"ahler 
transformation~(\ref{k-t}) is hard to describe. The reason is that the potential $K$ depends not only on Riemann 
surface data, but also on a specific choice of uniformization (Schottky e.g)~\cite{zo-tak}. Equivalently, to define $K$ 
one must pick a 3D space $V$ bounded by the Riemann surface $\S=\partial V$; $K$ depends on both $V$ and $\S$.
It is nevertheless 
possible to define wave functions that depend only on the surface data by using instead the Quillen norm~\cite{quill}. 
The redefinition of the wave function will be described in details later on; here we shall only state the key feature of such
redefinition: Quillen's~\cite{quill} definition, combined with the factorization theorem of Zograf~(see~\cite{tak-lap}) shows that
there exist a holomorphic function $H(\mu)$, nowhere zero in the interior of $T(\S)$, transforming under $M$ as 
\beq
H\rightarrow H'= {\det}^{c/2}(C\Omega + D) H \exp (-F), \qquad 
{\cal S}=\left(\begin{array}{ll} A & B \\ C& D \end{array} \right) \in Sp(2g).
\eeq{m1}
The constant $c$ is related to the Newton constant and AdS$_3$ radius by the classical 
(Brown-Henneaux) relation~\cite{BH} $c=3l/2G$ (we will consider only even integer $c$ to avoid problems with the 
definition of the square root). The $Sp(2g)$ matrix ${\cal S}$ describes the transformation of Abelian differentials under the 
mapping class group, while  $\Omega$ is the Abelian differentials' period matrix~\cite{far-kra}. 
Multiplying the wave function $\psi(\mu)$ by $H(\mu)$, we obtain a wave function $\Psi=H\psi$ whose
 transformation properties under the mapping class group depends only on Riemann surface data 
 and whose norm is the Quillen norm. 
Once quantization of boundary degrees of freedom is taken into account, physical  quantum gravity wave functions are thus {\em normalizable} (in the Quillen norm) entangled states 
\beq
\mathfrak{v}= v_\lambda \otimes u_{\lambda'} \otimes \Xi, \qquad \Xi = \sum_{IJ} N^{IJ}\Psi_I (\mu) \Psi_J (\bar\mu') .
\eeq{m2}
The vectors $v,u$ belong to irreducible representations of the Virasoro algebra with central charge $c=3l/2G$ and $\Xi$ 
transforms under $M$ as
\beq
\Xi \rightarrow \Xi'= {\det}^{c/2}(C\Omega(\mu) + D) {\det}^{c/2}(C\overline{\Omega}(\bar{\mu}') + D)\Xi.
\eeq{m3}

Eq.~(\ref{m3}) can be rephrased as saying that $\Xi$ transforms under the mapping class group as a power of the 
determinant bundle over $\S$; when $\S$ has one boundary this is the transformation property of 
expectation values of CFT primary 
operators over $\S$,
$\langle O \rangle_\Sigma$.  This can in fact be taken as a definition of such objects~\cite{seg,w8}. 
Notice that $\Xi$ depends holomorphically on the Teichm\"uller moduli 
$\mu, \bar{\mu}'$, so the correspondence with VEVs should read more precisely
\beq
\Xi |_{\mu'=\mu}\sim \langle O \rangle_\Sigma.
\eeq{m4}

Seen as a wave function, $\Xi$ is the (linear combination of) VEVs of  operators in a special class of CFTs. We will show 
later on that the conformal blocks contributing to $\Xi$ belong to CFTs that have no $\slc$ invariant
ground state, whose primary operators obey the Seiberg bound~\cite{seib} $\Delta> (c-1)/24$ and that 
possess a continuous spectrum of conformal weights. One  example of such theory is of course Liouville 
theory~\cite{seib,tesch-liu}, so the space of normalizable $\Xi$ states is not empty.

In fact, we will be faced with the opposite problem: the Hilbert space obtained by canonical quantization will turn out to be
too large. For one thing, Riemann surfaces of different genus give orthogonal subspaces. So, given
a CFT with a primary operator $O$, we end up with an infinity of states $\langle O\rangle_g$, indeed at least one for each 
genus, all associated with
the same representation of the Virasoro group. Moreover, if several CFTs satisfy our constraints and if a primary operator with 
the same conformal weight $\Delta$ exists in each of them, any linear combination 
$\Xi \sim \sum_{i\in CFTs} \alpha_i \langle O\rangle^i$ is a physical state with the same conformal weight. 
Both the former result and the latter  are incompatible 
with a reasonable holographic interpretation of 3D gravity and give rise to a large (in fact infinite) degeneracy of states 
at any energy. 

\subsection{Topological Transitions and Projections Over the Hilbert Space}

Such ``embarrassment of riches" has been noticed in a related context in~\cite{w8}. There, the Hilbert space resulting from 
quantization of Chern-Simons gravity on a closed Riemann surface was interpreted as the target space for CFTs partition
functions. We would like 
instead to interpret our space as ``the" Hilbert space of 3D quantum gravity on AdS. We thus need a means of
projecting the much too large space that 
we have obtained on a subspace that obeys some reasonable properties. To this end, we will study
topology changing amplitudes. These amplitudes were studied by Witten in 1989 for zero cosmological constant 3D 
gravity~\cite{w89}. We will show that even when $\Lambda<0$, pure gravity admits nonzero topology changing
amplitudes, that we will estimate using a (complex) saddle point approximation to the functional integral. 

The existence of such amplitudes is a bit puzzling because time evolution maps the Hilbert space $\mathfrak{H}_g$ obtained by 
canonical quantization on a surface of genus $g$ into itself. In the basis for $\mathfrak{H}_g$ that we defined earlier, which has
the form $v\otimes u \otimes  \varphi(\mu,\bar{\mu}')$, only the vectors $v,u$, belonging to  irreducible representations of 
the  Virasoro$\times$Virasoro algebra, transform under time evolution. 

One possible solution to this apparent contradiction is to {\em define} the correct physical space of the theory as that
which is
left invariant by the topological transition map. More precisely: a transition amplitude $P_{AB}$ from a state $A$ to $B$ 
obeys the composition property $P_{AB}=\sum_C P_{AC}P_{CB}$~\cite{w89}. It thus naturally defines a projection operator
on $\mathfrak{H}=\sum_g \mathfrak{H}_g$, whose image will be our definition of the physical Hilbert space. 

We will argue that the space defined by such projection is precisely the Hilbert space of Liouville field theory. We will conclude our study by analyzing some of the features of such theory as a potential dual of a gravity theory. We will
attempt to reconcile the continuous spectrum and no ground state features of Liouville theory with the behavior of a 3D AdS
gravity without states below the BTZ bound~\cite{BTZ}. We will also see how, as first pointed out in ref.~\cite{gkrs},
 the apparent discrepancy between the 
multiplicity of states in Liouville theory and the Bekenstein-Hawking entropy could be not a problem but instead 
an expected feature of pure gravity.

\subsection{Non-Geometric Projections}
The key feature that identifies quantum gravity wave functions with conformal blocks of CFTs with continuous spectrum is
that the moduli space of the theory is noncompact and has infinite volume; namely it is
$T(\S)\times T(\S)$ quotiented by only {\em one} copy of the mapping class group. 

It is conceivable that other non-geometric discrete gauge symmetries may appear at finite $c$, especially 
in the strong coupling regime 
$c=3l/2G \lesssim 1$. One simple possibility is to mod out each $T(\S)$ by the whole mapping class group.
The phase space in this case factorizes into two copies of the Riemann surface moduli space 
${\cal M}= T(\S)/M$. K\"ahler quantization on ${\cal M}$ produces normalizable wave functions which are
 conformal blocks of CFTs  with integer spectrum of dimensions {\em but without $SL(2,\mathbb{C})$ invariant vacuum for $c>1$.}
So, in spite of superficially resembling the factorized  CFT proposed as dual to pure gravity by Witten in 2007~\cite{w8}, the
CFTs appearing in quantizing ${\cal M}$ are different. In particular, the multiplicity of states predicted by such CFTs is still
parametrically smaller than the exponential of the Bekenstein-Hawking entropy. 

A more interesting possibility arises at $c<1$. This is a deep quantum regime without semiclassical parallel. For $c<1$, the
conformal blocks of unitary CFT on a torus are invariant under a normal, finite index subgroup of the modular group~\cite{bant}. Moreover, the $\slc$-invariant vacuum is in this case normalizable. If such kernel of the mapping class group exists also
on higher-genus Riemann surfaces,  then it makes sense to
quotient $T(\S)$ by it. If such kernel ${\cal N}$ is also as for the torus normal and finite-index, then the quotient space 
$T(\S)/{\cal N}$ is compact and K\"ahler quantization will produce conformal blocks with discrete spectrum. So, 
by imposing ${\cal N}$ as a gauge symmetry of 3D pure gravity, one would select the ``right" dual theories,  namely 
unitary minimal models. 
Some of these models have already appeared as possible duals of pure AdS$_3$ gravity in~\cite{ising}.

\subsection{Chiral Gravity}
The Chern-Simons formulation of gravity becomes singular when one of the levels $k, k'$ vanishes. The second-order 
 metric formulation is instead a potentially consistent, ghost-free theory of gravity, called 
 {\em chiral gravity}~\cite{stro1,stro2,stro3}. It is in some sense a ``square root'' of pure gravity. A canonical quantization 
 of pure gravity, particularly tailored to chiral gravity has been performed in~\cite{maloney}. Ref.~\cite{maloney} finds a 
 compact phase space with a discrete, finite multiplicity spectrum of black-hole microstates.  We thank A. Maloney for
 making his paper available to us prior to publication and for coordinating with us the joint release of our papers. 

\subsection{Plan of the Paper}
In section 2 we will review the connection between the second-order, metric formulation of pure AdS$_3$ gravity and the 
Chern-Simons formulation. We will show in particular that possible differences in the phase space of the
two theories do not contradict the fact that they are classically equivalent as theories of gravity and that Chern-Simons 
gravity contains a sector of its phase space identical with the phase space of gravity in the metric formulation.

In section 3 we will study in details the phase space of one of the two $SL(2,\mathbb{R})$ components of pure gravity 
Chern-Simons for the topology $\S=$ a one-boundary torus. We will show how the phase space decomposes into simple 
components, each one endowed with a well-understood symplectic form. This will allow us to relate
directly the moduli space of flat $SL(2,\mathbb{R})$ connections to Teichm\"uller space. 

In section 4 we will extend the results of section 3 to generic one-boundary Riemann surfaces and quantize the phase
space using holomorphic polarizations and K\"ahler quantization. We will show how the
natural norm that defines the Hilbert space of quantum gravity is obtained by using the Quillen norm. We will also show how the
 holomorphic quantization that we defined is related to other possible choices of polarization and
other quantizations.

Section 5 is the core of the paper. There we will impose normalizability and modular invariance of state vectors and we will
obtain the Hilbert space of canonically quantized AdS$_3$. Section 5 will also discuss
how to relate the K\"ahler potential of surfaces with boundaries to the potential of surfaces with punctures and it will also
relate the Hilbert space to CFTs.

In section 6, topological transition amplitudes will be used to project the Hilbert space found in section 5 to a subspace 
that can be identified with a holographic dual CFT. 

Various comments and observations on such proposed dual and its properties will conclude  the paper in section 7. 
Section 7 will also discuss in more details the possibility of obtaining different dual CFTs by quantizing quotients of the
phase space $[T(\S)\times T(\S)]/M$, especially in the strong gravity case $c=3l/2G<1$. 

Two appendices detail the computation of the volume of the moduli space for the torus with one geodesic boundary and a 
change of coordinates between two common coordinate systems for \tch\ space, also in the case of the 
torus with one geodesic boundary. 

\section{Chern-Simons vs Metric Formulation of Gravity}
The canonical formalism constrains the topology of an asymptotically AdS$_3$ space-time to be 
$\S \times \mathbb{R}$, where 
$\S$ is a Riemann surface with a single boundary; the boundary is topologically a circle
$\partial \S =S_1$. Once a metric and its conformal structure are introduced, the space becomes metrically complete and
$\partial \S$ becomes the conformal boundary. 
Using coordinates 
$(r,x^+=t/l+\theta,x^-=-t/l+\theta)$, the asymptotic metric is
\beq
d s^2 \approx l^2 \left(d r^2  + e^{2 r} d x^+ d x^-\right), \qquad \qquad r
\rightarrow \infty.
\eeq{m5}
The Einstein equations $R_{\mu\nu}-(1/2)g_{\mu\nu} R= (1/l^2)g_{\mu\nu}$ with boundary conditions~(\ref{m5}) can be solved for an arbitrary surface $\S$. The
geometric and causal properties of such solutions have been studied in~\cite{brill,barbot}. An explicit (though
not complete) metric, which gives an explicit parametrization of the phase space of pure 3D AdS was given in~\cite{ks}. 
That paper used the results of~\cite{monc} and generalized them to spaces with boundaries. With or without boundaries,
the solution to Einstein's equations on $\S \times \mathbb{R}$ is parametrized by a metric that can be written, locally in a holomorphic coordinate chart $(z,\bar{z})$, as 
\bea
ds^2 &=& -d\tau^2 +\cos^2 \tau e^{2\phi} dz d\bar{z} + \sin \tau \cos \tau [q dz^2 + \bar{q} d\bar{z}^2 ] + \sin^2 \tau e^{-2\phi} q\bar{q} dz
d\bar{z}, \label{m6} \\
4\partial\bar{\partial} \phi &=& e^{2\phi} - e^{-2\phi} q\bar{q}.
\eea{m7}
In these equations, the 3-manifold $\S \times \mathbb{R}$ is foliated by maximal surfaces (with zero mean curvature). The
Hamiltonian and momentum constraints --that is the Gauss-Codazzi equations-- impose that $q$ is a holomorphic quadratic differential~\cite{ks}. The phase space defined by this parametrization is naturally identified with $T_* T(\S)$, the cotangent bundle 
over the Teichm\"uller space of $\S$. This was proven for compact surfaces in~\cite{monc} and extended to surfaces
with one boundary (indeed to the universal Teichm\"uller space) in~\cite{ks}.  That the phase space is $T_* T(\S)$ is evident already
from the parametrization in eqs.~(\ref{m6},\ref{m7}), since the data needed to define it are a complex structure, defined
by the coordinate system ($z,\bar{z}$), and 
a holomorphic quadratic differential. What is far from evident is that the phase space is also $T(\S)\times T(\S)$. More
precisely, that there exists a one-to-one smooth map 
\beq
T_*T(\S)\leftrightarrow T(\S)\times T(\S).
\eeq{m8}
The existence of such bijection was proven for compact manifolds $\S$ in~\cite{mess}; 
the proof was extended to manifolds
with boundary in~\cite{ks}. 

We present here a heuristic justification of such factorization, using formulas that will become useful when studying topology-changing amplitudes.\footnote{This argument was suggested in conversation to one of us (M.P.) by A. Maloney.} 
The idea is that in Euclidean signature and with topology $\S\times \mathbb{R}$, a solution of Einstein's equations with
negative cosmological constant 
is specified by giving data on  {\em two} conformal boundaries: the ``initial'' one at $r\rightarrow -\infty$ and the ``final" 
one at $r\rightarrow +\infty$. The data are in each case 2D 
metrics modulo local diffeomorphisms and local Weyl rescalings, i.e. points in the Teichm\"uller space of $\S\times \S$. 
The role of global diffeomorphisms is subtle because the topology of the 3D space is restricted to 
$\S \times \mathbb{R}$. So 2D global diffeomorphisms cannot act independently on the two Riemann surfaces: only a 
diagonal action of the mapping class group $M$ is permitted. This argument, which suggests (correctly) that the phase 
space of gravity is $[T(\S)\times T(\S)]/M$ will be discussed in more detail in section 5. If, for the time being, we mod out 
only by 2D diffeomorphisms connected to the identity we obtain a product phase space $T(\S)\times T(\S)$.

The equivalence of moduli spaces 
can be seen explicitly considering a special point in $T_*T(\S)$ as we will show now. 
The analytic continuation of eq.~(\ref{m6}) to Euclidean signature is effected by the substitutions $\tau\rightarrow ir$, $q\rightarrow  iq$, $\bar{q}\rightarrow i\bar{q}$ [{\em sic}]. The Euclidean metric is thus
\beq
ds^2_E= dr^2 +\cosh^2 r e^{2\phi} dz d\bar{z} + \sinh r \cosh r [q dz^2 + \bar{q} d\bar{z}^2 ] +\sinh^2 r e^{-2\phi} q\bar{q} dz d\bar{z}.
\eeq{m9}
Consider now a small deformation of a point $q=0,(z,\bar{z})$ in $T_* T(\S)$. The deformation changes 
$z\rightarrow z+ w(z,\bar{z})$ and makes $q$ nonzero. Thanks to a local Weyl transformation $\delta \phi$ such that 
$\partial w + \bar{\partial} \bar{w} + 2 \delta \phi =0$, the deformed metric becomes
\beq
ds'_E= ds_E^2 + \cosh r e^{2\phi} [(\partial \bar{w} \cosh r + \bar{q} e^{-2\phi} \sinh r ) dz^2 + (\bar{\partial} w \cosh r 
+ q e^{-2\phi} \sinh r) d\bar{z}^2] .
\eeq{m10}
So, the``initial" surface $r\rightarrow -\infty$ depends only on the linear combination $\mu_+= \bar{\partial}w + e^{-2\phi}q$,
while the ``final" surface depends only on $\mu_-=  \partial\bar{w} - e^{-2\phi}\bar{q}$. We thus identify $\mu_\pm$ with the
coordinates of factorized Teichm\"uller spaces. 
\subsection{Chern-Simons Moduli Space}
The moduli space of $SL(2,\mathbb{R}) \times SL(2,\mathbb{R})$ Chern-Simons theory 
on $\S \times \mathbb{R}$ factorizes naturally
 into a product of two moduli spaces of flat 2D $SL(2,\mathbb{R})$ connections.  
Each contains several components of different Euler class. The component with maximum Euler class ($2g-2$ for a
closed, genus-$g$ Riemann surface) is Teichm\"uller space and its natural symplectic form is the 
Weil-Petersson form~\cite{gold}. The natural 
symplectic form of $\slr$ connections used in~\cite{gold,hitch} is that which is induced by the CS action. 
 An argument showing explicitly the equivalence of moduli spaces was given in~\cite{V}. It follows from 
 parametrization~(\ref{m1a}), since $e^\varphi \neq 0$ and
 the flatness conditions~(\ref{constr}) imply that $\mu(z)$ is a holomorphic quadratic differential. The possibility of 
 parametrizing flat $\slr$ connections of maximum Euler class as in~(\ref{m1a}) in turns follows from the results of Hitchin 
 (see section 11 of ref.~\cite{hitch}). In fact~(\ref{m1a}) is essentially the same parametrization
used in~\cite{hitch}. 

We will give a very explicit construction of the CS symplectic form in the special case that $\S$ is the torus with one 
boundary in the next section. After that, section 4 will describe in details how to construct the flat $\slr \times \slr$ 
connections and symplectic forms out of the 3D metric data. 

Here we conclude with only a few remarks. The first one is that in
 spite of being topologically distinct from other components of the $F|_\S=0$ moduli space, the Teichm\"uller component is connected to other components, since Teichm\"uller space has a boundary. From the point of view of 
the geometry of the surface $\S$, this property is due to the possibility of changing the genus of $\S$ by pinching a handle. 
While the pinch geometry is singular, the singularity is relatively mild, so a generic wave function can in principle spread to 
other components of the moduli space. 

The second is that the CS formulation of pure gravity does not share a problem that exists instead in the theory with zero
cosmological constant~\cite{matsch}. The problem arises when considering two flat connections connected 
by ``small" gauge transformations, i.e. transformations connected to the identity. These may have to pass through ``walls" of singular configurations where  the  zweibein $e^\pm$ is not invertible. In this case it is no longer true that
the gauge transformation is equivalent to a diffeomorphism. So, the two configurations would be equivalent in the  CS
theory but not in the metric theory.  But this cannot happen in AdS$_3$ because two flat 
connections separated by such wall necessarily belong to spaces with different Euler class. 
The relation between CS and geometrical
gravity was investigated in depth in~\cite{ks} and, in the case of compact $\S$,  also in~\cite{meus}.

Finally, we must point out that in the ``constrain first" approach one never encounters the main difficulty found in the canonical quantization of $\Lambda=0$ pure 3D gravity. In the latter case, one can find an exact description of the
wave function of quantum gravity because the local frame $e^a$ and the spin connection $\omega^a$ are canonically
conjugate variables~\cite{w}. By choosing $\omega^a$ as coordinate, $e^a$ is represented canonically as $-i\delta/\delta \omega^a$.
When applied to a physical wave function, neither the operator $e^a$ nor  $e=\det e^a$ have any positivity property. In the 
functional integral approach this is even clearer (though a bit more heuristic): the integral 
$\int [de d\omega] \exp(-S_{CS}) O(\omega)$ can be done exactly, when the operator $O$ depends only on $\omega$. 
In such case, by integrating first on $[de]$ one gets a functional delta 
function that forces the connection $\omega$ to be flat: 
$\int [de d\omega] \exp(-S_{CS}) O(\omega)\sim \int [d\omega] \delta[d\omega +\omega\wedge \omega ] O(\omega)$. 
This nice result depends crucially on integrating over all values of $e^a$, without imposing any 
positivity constraint~\cite{w,w89}.  In our case instead, we do satisfy positivity constraints on the metric precisely by
reducing the phase space of flat $\slr$ connections to its \tch\ component.

\section{Example: Torus with a Geodesic Boundary}\label{torus}

We argued that the Chern-Simons formulation of pure three-dimensional gravity is equivalent to gravity in the
metric formulation.
The relevant phase space is constructed from two copies of the moduli space of flat  $SL(2, \mathbb{R})$ connections. 
We will show explicitly in this section that the $\slr$ moduli space for the torus with one geodesic boundary, $T_{(1,1)}$,  
contains the 
$T_{(1,1)}$ \tch\ space and that the symplectic form induced by the CS action is the Weil-Petersson form of $T(T_{(1,1)})$. 

On the spacetime manifold $M=T_{(1,1)}\times \mathbb{R}$ the general form of the solution of the Chern-Simons 
equations of motion is $\tilde{A}= g^{-1} \tilde{d}g + g^{-1} H g$ with $g$  a single-valued function from $T_{(1,1)}$ to 
$SL(2, \mathbb{R})$ and $H$ a flat connection specified by holonomies around non-contractible cycles of $T_{(1,1)}$.
This led us to action (\ref{action_eff}), which can be decomposed into three parts
\bea
I_{CS} &=& I_1 + I_2 + I_3 , \nonumber\\
I_1 &=& \frac{k}{4 \p}\int_{\mathbb{R}}{dt \int_{\pd T_{(1,1)}}{ \mathrm{Tr} \left( g^{-1} \td{d} g g^{-1} \dot{g} \right) }} + \frac{k}{12 \p}\int_{M}{\mathrm{Tr} \left( g^{-1} {d} g \right)^3 } , \nonumber \\
I_2 &=&  -\frac{k}{2 \p}\int_{\mathbb{R}}{dt\int_{\pd T_{(1,1)}}{\mathrm{Tr}\left( g^{-1} \dot{g} H \right)}} , \nonumber \\
I_3 &=& - \frac{k}{4 \p} \int_{\mathbb{R}}{dt\int_{\S}{\mathrm{Tr} \left( H \wedge \dot{H} \right)}} .
\eea{action_3_1}
We can define two boundaries here; one is conformal boundary which comes from Brown-Henneaux boundary 
condition and the other is a geodesic boundary 
whose length is a function of the black hole mass $M$ and angular momentum $J$.
The spatial geometry is thus an annulus attached to the Riemann surface  $T_{(1,1)}$ (see figure~(\ref{collar}) in 
section 4).
The construction of Einstein metrics on such space is studied, for example, in~\cite{brill}.

Once a geometry is given,  one may want to identify the action $I_1$ with a WZW action defined on an annulus, but this is not a correct approach, because $g$ is a single valued function.
Rather, we must identify the action $I_1 + I_2$ as a WZW action defined on a disk with a source $-k H$~\cite{MS}.
Its symplectic form is
\beq
\frac{k}{4 \pi} \oint \mathrm{Tr} \left[ \left( g^{-1} \delta g \right) \frac{d}{d\theta} \left( g^{-1} \delta g \right) \right] -\frac{k}{2 \pi} \oint \mathrm{Tr} \left[ \left(g^{-1} \delta g \right) H \right].
\eeq{symp_I_12}
The action $I_1 + I_2$ is invariant under $ g \rightarrow U(\theta) g V(t)$.
Since the invariance under $U(\theta)$ implies that a physical state should fall into a representation of the loop group and $V(t)$ should commute with $H$, the Hilbert space will be a lowest-weight representation of the Ka\v{c}-Moody algebra characterized by $H$~\cite{p&s}.

A theory with action $I_1 + I_2$ is well-undersood and will be studied in more details in the next section, but action $I_3$ requires more care.
The symplectic form is easily obtained from action $I_3$ following a general recipe for a first order theory~\cite{w_bos},
\beq
\w_{\alpha \beta} = -\frac{k}{4 \pi} \int \mathrm{Tr} \left[ \delta_{\alpha}H \wedge \delta_{\beta}H \right],
\eeq{symp_H}
where $\delta_\alpha$ and $\delta_\beta$ label variations along two coordinates of the moduli space of flat $\slr$
connections on $T_{(1,1)}$.
The holonomy group of a genus $g \geq 1$ Riemann surface $\S_{(g,1)}$ with a geodesic boundary is generated by $\left(A_1 , \cdots , A_g , B_1 , \cdots , B_g ; D \right)$, where $A_i$ and $B_j$ generate holonomies $W_{A_i}$ and $W_{B_j}$ around the fundamental cycles of genus $g$ surface and $D$ generates the holonomy $W_D$ around the geodesic boundary. Together, $W_{A_i},W_{B_i},W_D$ define the flat connection up to a global gauge transformation. 
The holonomies should be hyperbolic elements of $\slr$ and they are constrained by the relation
\beq
W^{-1}_{B_g} W^{-1}_{A_g} W_{B_g} W_{A_g} \cdots W^{-1}_{B_1} W^{-1}_{A_1} W_{B_1} W_{A_1} = W_D .
\eeq{hol_const}
Also, holonomies are identified modulo conjugation by an  element $\slr$
\beq
W_i \sim W'_i \iff \exists \; G\in  \slr \mbox{ such that } W_i=GW'_iG^{-1} \; \forall i, \qquad 
W_i=W_{A_i},W_{B_i} \mbox{ or } W_{D}.
\eeq{conj}

So the moduli space of flat $\slr$ connections on $\Sigma_{(g,1)}$ will be 
isomorphic to a $6g - 6 + 3$ dimensional real space.
This is true if the generator $D$ has three free parameters.
In our case those parameters correspond to the position of the geodesic boundary in the two dimensional space and its length.
But the length is related to a black hole mass $M$ and angular momentum $J$, so it becomes a constant rather than a free parameter once we fix the black hole parameters.
The other two parameters can be gauged away using some part of modular transformation, which in turn give us
 $6g - 4$ real  parameters; $6g$ comes from $2g$ generators around the fundamental cycles, $-3$ comes from the 
 constraint (\ref{hol_const}) and $-1$ comes from identification of holonomies under an overall $\slr$  transformation that
commutes with $D$.

Notice that we fixed the holonomy around the geodesic boundary $W_D$ as a constant for every time slice.
To check the consistency of this choice let us write it as
\beq
W_D = \mathrm{Tr} \left[  \mathcal{P} \exp \left( \oint A \right) \right] ,
\eeq{W_D}
where $A$ is the $\slr$ connection on the surface.
Using the equations of motion in the three dimensional theory and the fact that $D W_D =0$, the time derivative $W_D$ is
\bea
\frac{d}{dt} W_D &=& \oint \mathrm{Tr} \left[\mathcal{P} \exp \left( \oint A \right) \dot{A} \right] ~ = \oint \mathrm{Tr} \left[ \mathcal{P} \exp \left( \oint A \right) D A \right], \nonumber \\
 &=& \oint \mathrm{Tr} d \left[ \mathcal{P} \exp \left(\oint A \right) A \right] = 0.
\eea{dtdW}

The symplectic form (\ref{symp_H}) can be written as a one-dimensional integral using the fact that the variation of $H$ with respect to moduli parameters is locally a pure gauge, i.e. $ \delta_{\alpha} H = D \w_{\alpha}$ with $\w_\a = \d_\a W W^{-1}$, 
\bea
\w_{\a \b} &=& -\frac{k}{4 \pi} \int \mathrm{Tr} \left[ D \w_\a \wedge  D \w_\b \right] = -\frac{k}{4 \p} \int d \mathrm{Tr} \left[ \w_\a \wedge D \w_\b \right]  \nonumber \\
 &=& - \frac{k}{4 \p} \oint_{hc} \mathrm{Tr} \left[ \w_\a \wedge \d_\b H \right] .
\eea{symp}
The contour integral in the last expression is defined as in subsection 1.2 by cutting the surface along a 
homology cycle $hc$ in such a way as to obtain topologically a disk with a hole as in figure~(\ref{cutsurface}). The explicit 
mapping of specific homotopy cycles depends on the surface. In our case the result is illustrated in figure~(\ref{square}). 

In the rest of this section we specialize our computation to the torus with a geodesic boundary and give a detailed discussion of the phase space obtained from action $I_3$, including an explicit calculation of the symplectic form~(\ref{symp}).
More general Riemann surfaces will be discussed in section 4.
 
\subsection{Parametrization}

Let us start by finding an explicit parametrization for the holonomy generators.
The holonomy group of a one-holed torus is generated by three elements; $A$ and $B$ along the fundamental cycles of torus and $D$ along the geodesic boundary; they are constrained by a relation
\beq
e^{-B} e^{-A} e^B e^A = - W_D .
\eeq{constraint}
We argued that $W_D$ can be chosen as a constant matrix and a convenient choice is $W_D = \mathrm{diag} ( e^{\lam}, e^{-\lam} )$, where $2\lam$ is the length of the geodesic boundary.
The $(-)$ sign in the right hand side of eq.~(\ref{constraint}) is essential to define holonomies around a torus with a hole, since a $(+)$ sign will 
give us instead the holonomies of a sphere with three holes.
To see this let $e^A$ and $e^B$ be
\begin{equation}
e^A = \begin{pmatrix} \xi & 0 \cr 0 & \frac{1}{\xi} \end{pmatrix} ,\quad e^B = \begin{pmatrix} a & b \cr c & d \end{pmatrix} ,
\end{equation}
with $ad-bc=1$.
The matrix $e^A$ defines an isometry of the upper half plane with hyperbolic metric $ds^2= dzd\bar{z}/\Im z$; 
the geodesic joining the fixed points of the isometry (the ``axis" of the transformation) is in this case the imaginary axis of the $z$ plane.
The axes of $e^A$ and $e^B$ should intersect  each other to be holonomies of a torus with a hole, so $e^B$ should have one positive fixed point and one negative fixed point.
Since the fixed points of $e^B$ are 
\begin{equation}
z = \frac{(a - d) \pm \sqrt{(a - d)^2 + 4 bc}}{2c},
\end{equation}
the product $bc$ should be positive for $e^B$ to have fixed points of both signs.
A simple computation then shows that the trace of $e^{-B} e^{-A} e^B e^A$ is $2 + bc \left( 2 - {\xi}^2 - \frac{1}{{\xi}^2} \right)<2$. Since $\Tr e^{-B} e^{-A} e^B e^A$ is hyperbolic, it is either larger than 2 or less than $-2$; this
implies that $(-)$ is the correct sign to choose.

To find a parametrization of the moduli space of flat $\slr$ connections, let us write next $A$ as
\bea
A &=& \a \left( \a_a T^a \right) ~\equiv~ \a \hat{A} , \nonumber \\
T^1 &=& \begin{pmatrix} 0 & 1 \cr 1 & 0 \end{pmatrix}, \quad T^2 ~=~ \begin{pmatrix} 0 & -1 \cr 1 & 0 \end{pmatrix}, \quad T^3 ~=~ \begin{pmatrix} 1 & 0 \cr 0 & -1\end{pmatrix} ,
\eea{matA} 
where $\a^2_1 - \a^2_2 + \a^2_3 = 1$.
When we right-multiply both  sides of equation~(\ref{constraint})  by $e^{-A}$  we have
\beq
-W_D e^{-A} = \exp \left( - e^{-B} A e^B \right) \equiv \exp \left( \a \hat{A}^R \right),
\eeq{constraint2}
where $\hat{A}^R = \a^R_a T^a$ can be interpreted as a \emph{rotated} unit vector of $\hat{A}$ under an element of $\slr$.
By equating each term of the constraint (\ref{constraint2}), we have
\bea
\a^R_1 &=& - \a_1 \cosh \lam - \a_2 \sinh \lam , \nonumber \\
\a^R_2 &=& - \a_1 \sinh \lam - \a_2 \cosh \lam , \nonumber \\
\cosh\a \pm \a^R_3 \sinh \a &=& - e^{\mp \lam } \left( \cosh \a \pm \a_3 \sinh \a \right) .
\eea{const_compo}
The first two equations come from the off-diagonal terms of constraint (\ref{constraint2}), and imply that $\a^R_3 = \pm \a_3$.
But only the negative sign is compatible with the last equation, which comes from diagonal terms and gives us
\bea
\left( \a^R_1 , \a^R_2 , \a^R_3 \right) &=&  \left( -\a_1 \cosh \lam - \a_2 \sinh \lam, -\a_1 \sinh \lam - \a_2 \cosh \lam, -\a_3 \right) , \nonumber \\
\a_3 &=& \coth \left( \lam / 2 \right) \coth \a .
\eea{roatated}
The first consequence of this result is that $\a_3$ is not a independent parameter: it is completely fixed by the length of the boundary $\lam$ and the \emph{length} of the generator $A$.
The second consequence is that, apart from an overall negative sign, $\a^R$ looks like a transformation 
of $\a$ under a boost along $3$-direction.
The sign of $\a_1$ and $\a_3$ can be inverted by the action of $\exp \left( \pm \frac{\p}{2} T^2 \right) = \mp T^2$.
So we have to find a transformation generated by $T^1$ to have $(\a_2, -\a_3 ) \rightarrow -(\a_2 , \a_3)$.\footnote{We cannot use a transformation generated by $T^3$ because it will give another boost  along the direction $3$.}
A little algebra shows that we can explicitly find such a transformation:
\bea
e^B &=& e^{\frac{\p}{2} T^2 } e^{\r T^1} e^{-\frac{\lam}{2} T^3} , \nonumber \\
\a_2 &=&  -\a_3 \tanh \r , \nonumber \\
\a_1 &=& \pm \sqrt{ 1 - \mathrm{coth}^2 \a  ~ \mathrm{coth}^2 \left( \lam / 2 \right) ~ \mathrm{sech}^2 \r } .
\eea{e^B}
The $\a_2$ parameter is chosen to obey $(\a_2, -\a_3 ) \rightarrow -(\a_2 , \a_3)$ under the action of $ e^{\r T^1}$ while
 $\a_1$ is fixed by normalizing the vector $(\a_1,\a_2,\a_3)$ to unit lenght.
However parametrization (\ref{e^B}) is not the only choice of $e^B$, because $A$ is invariant under the action of $e^{\b \hat{A}}$ for any $\b\in \mathbb{R}$.
So the most general form of $e^B$ is
\beq
e^B = e^{ \b \hat{A} } e^{\frac{\p}{2} T^2 } e^{\r T^1} e^{-\frac{\lam}{2} T^3}.
\eeq{paraB}

At first glance it look like that we have three ``parameters'' $\a$, $\b$ and $\r$.
But the moduli space of $\slr$ holonomies of the one-holed torus with fixed boundary is two dimensional, so one of the
parameters must be redundant.
To figure out the correct moduli parameters, one can  notice that $\a$ and $\b$ correspond to the length of the 
hyperbolic vectors $A$ and $B$, and $\r$ is related to their direction.
Under a transformation of $\slr$ which leaves $W_D$ invariant, the direction of vectors may change, but their length 
should be invariant. To be more precise let us consider the action of $T = e^{\rho T^3}$.
The constraint equation transforms as
\beq
T : K^{-1}  e^{-\a \hat{A} }  K e^{\a \hat{A} } = -W_D \rightarrow \left(K^{\prime} \right)^{-1}  e^{-\a \hat{A}^{\prime} }  K^{\prime} e^{\a \hat{A}^{\prime} } = -W_D ,
\eeq{constraint3}
where $ K = e^{\frac{\p}{2} T^2 } e^{\r T^1} e^{-\frac{\lam}{2} T^3} $, and the prime denotes transformed objects under the action of $T$.
The form of the transformed constraint equation is the same as that of the original one, so $\hat{A}^{\prime}$ transforms under $K^{\prime}$ in the same manner as $\hat{A}$ transforms under $K$.
Also all the component associated with the direction ``3'' of primed objects are the same as their counterparts of unprimed objects; so, the net effect of the action of $T$ is to rotate $A$ in the $(12)$-plane and change $\r$ to $\r^{\prime}$, the latter being related to the direction of $A^{\prime}$.
Therefore, $\r$ should be interpreted as a redundant parameter, and all physical quantities should be independent of it.

\subsection{Phase Space}

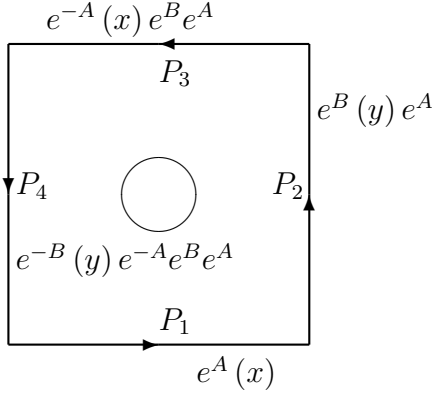
\begin{figure}[h]
\begin{center}
\setlength{\unitlength}{1mm}
\begin{picture}(70,50)

\put(35,25){\circle{10}}

\thicklines
\put(15,5){\vector(1,0){20}}
\put(35,5){\line(1,0){20}}
\put(55,5){\vector(0,1){20}}
\put(55,25){\line(0,1){20}}
\put(55,45){\vector(-1,0){20}}
\put(35,45){\line(-1,0){20}}
\put(15,45){\vector(0,-1){20}}
\put(15,25){\line(0,-1){20}}

\put(56,35){$e^B \left( y \right) e^A$}
\put(20,47){$ e^{-A} \left( x \right)  e^B e^A$}
\put(40,0){$e^A \left( x \right)$}
\put(16,15){$e^{-B} \left( y \right)  e^{-A} e^B e^A$}

\put(50,25){$P_2$}
\put(35,40){$P_3$}
\put(35,7){$P_1$}
\put(16,25){$P_4$}

\end{picture}
\end{center}
\caption{A cut surface $T_{(1,1)}^C$ giving a one-hole torus upon identification of the square boundary segments. 
We define the holonomy $W(z)$ along each edge of the square so that we have $e^{-B} e^{-A} e^B e^A$ when 
we complete a cycle. }
\label{square}
\end{figure}

After having found an explicit parametrization of holonomies, we are finally in a position to study the phase space of the action $I_3$.
Let us start with computing the symplectic form (\ref{symp}):
\beq
\w_{\a \b} = - \frac{k}{4 \p} \oint_{hc} \mathrm{Tr} \left[ \w_\a \wedge \d_\b H \right] .
\eeq{symp2}
Here the moduli parameters $\a$ and $\b$ can be identified with those in the parametrization of $A$ and $B$ given earlier.
A one-holed torus  can be represented as a square with a hole in its center.
On the cut surface $T_{(1,1)}^C$, shown in figure~(\ref{square}),  a gauge field can be represented {\em everywhere} as  
$H=dW(z)W^{-1}(z)$ for an appropriate definition of $W(z)\in \slr$, $z\in T_{(1,1)}^C$.  We define the holonomy $W(z)$ 
along  each edge of the square as indicated in figure~(\ref{square}). 

Along the path $P_1$ we define $W(x)=e^{f(x) A}$ with $f(x)$ such that $f(0)=0$ and $f(1)=1$.
The gauge connection along this path is $ d f(x) A$ and it vanishes when we vary it with respect to $\b$.
So the symplectic form gets no contribution from the path $P_1$.
Path $P_3$ does not contribute to the symplectic form for the same reason.

We want to define $W(z)$  along the path $P_2$ in such a way that it is $e^B$ at the end of the path and its variations with
respect to both $\a$ and $\b$ are easy to compute. 
Let us break the path into four subpaths $P_{2a}$, $P_{2b}$, $P_{2c}$ and $P_{2d}$ to perform the computation.
On each of the subpaths we define monotonically increasing functions $f_{2a}(y)$, $f_{2b}(y)$, $f_{2c}(y)$ and $f_{2d}(y)$ 
such that
\bea
f_{2a} (0) = f_{2b} (1/4) = f_{2c} (2/4) = f_{2d} (3/4) = 0, \nonumber \\
f_{2a}(1/4) = f_{2b} (2/4) = f_{2c}(3/4) = f_{2d}(1) = 1 ,
\eea{func_y}
so that the holonomy along the path $P_2$ is given by
\beq
W=
\begin{cases}
e^{f_{2a} (y)  \left( -\frac{\lam}{2} T^3 \right) } e^A & \quad 0 \leq y \leq 1/4 , \\
e^{f_{2b} (y) \r T^1 } e^{    -\frac{\lam}{2} T^3  } e^A & \quad 1/4 \leq y \leq 2/4 , \\
e^{f_{2c} (y) \frac{\p}{2} T^2} e^{ \r T^1 } e^{ -\frac{\lam}{2} } e^A &  \quad 2/4 \leq y \leq 3/4 , \\
e^{f_{2d} (y) \b \hat{A}} e^{\frac{\p}{2} T^2} e^{ \r T^1 } e^{ -\frac{\lam}{2} T^3  } e^A & \quad 3/4 \leq y \leq 1 .
\end{cases}
\eeq{W_B}
The subpaths $P_{2a}$, $P_{2b}$ and $P_{2c}$ give no contribution to the symplectic form because the gauge connections are independent of the moduli parameter $\beta$ along those subpaths.
On subpath $P_{2d}$, on the other hand, the gauge connection is $ d f_{2d} (y ) \b \hat{A}$, so $\d_\b H = d f_{2d}(y) \hat{A}$.
To compute $\w_\a$ on $P_{2d}$, we can use the fact that the variation of $\psi = A e^h B$ is given by $\d \psi \psi^{-1} = A \left[ \int^1_0 ds ~ e^{sh} \d h e^{-sh} \right]  A^{-1}$ if $\d h$ does not commute with $h$.
So $\w_\a$ on $P_{2d}$ is
\beq
\w_\a = \int^1_0 ds ~ e^{s \chi \b \hat{A} } \d_\a \left( \chi \b \hat{A} \right) e^{- s \chi \b \hat{A}}  + \int^1_0 ds ~ e^{\chi \b \hat{A} } K  e^{s \a \hat{A}} \d_\a \left( \chi \b \hat{A} \right) e^{-s \a \hat{A}} K^{-1} e^{- \chi \b \hat{A}} , 
\eeq{w_a}
where $\chi = f_{2d} (y)$.
Using these expressions for $\d_\b H $ and $\w_\a$, we can compute all nonzero contribution to the symplectic form (\ref{symp2}) along the path $P_2$:
\beq
\w^{P_2}_{\a \b} = -\frac{k}{4 \p} \int^1_0 ds ~ \mathrm{Tr} \left[ \left( e^{s \a \hat{A} } \d_\a \left( \a \hat{A} \right) e^{- s \a \hat{A}} \right) \left( K^{-1} \hat{A} K \right) \right].
\eeq{w_P2}
A similar calculation shows that the contribution from the path $P_4$ is
\beq
\w^{P_4}_{\a \b} = -\frac{k}{4\p} \mathrm{Tr} \left[ \d_\a \left( \a \hat{A}  \right) \hat{A} \right] +\frac{k}{4 \p} \int^1_0 ds ~ \mathrm{Tr} \left[ \left( e^{s \a \hat{A} } \d_\a \left( \a \hat{A} \right) e^{- s \a \hat{A}} \right) \left( K^{-1} \hat{A} K \right) \right].
\eeq{w_P4}
Therefore the total contribution is 
\beq
\w_{\a \b} = -\frac{k}{4\p} \mathrm{Tr} \left[ \d_\a \left( \a \hat{A} \right) \hat{A} \right] = -\frac{k}{2 \p}.
\eeq{symp3}
This means that our parametrization defines canonical variables over the phase space.
This is already a nice result but we can do better than this by transforming to  Fricke-Klein  coordinates~\cite{fk}.

Fricke-Klein  coordinates are defined in terms of traces of the holonomies $e^A$, $e^B$ and $e^Ae^B$; they are related to our parametrization by 
\bea
x&=&\mathrm{Tr} \left( e^A \right)  = 2 \cosh \a , \nonumber \\
y&=&\mathrm{Tr} \left( e^B \right) = 2 \sinh  \frac{\lam}{2}  \left[ \cosh \b \sinh  \frac{\r}{2}  + \a_1 \sinh \b \cosh  \frac{\r}{ 2 }  \right] ,  \nonumber \\
z&=&\mathrm{Tr} \left( e^A e^B \right) = 2 \sinh  \frac{\lam}{ 2} \left[ \cosh \left(\a + \b \right) \sinh  \frac{\r}{ 2 }  + \a_1 \sinh \left( \a + \b \right) \cosh \frac{\r}{ 2}  \right].
\eea{fricke}
Then a straightforward computation shows that
\beq
x^2 + y^2 + z^2 = xyz + 2 \left( 1 - \cosh \lam \right),
\eeq{region}
which is the same constraint equation that defines the  Teichm\"uller space of a one-hole torus~\cite{keen}.
Moreover, we can express the symplectic form~(\ref{symp3}) in terms of $x$, $y$, and $z$.
Using the intermediate coordinates
\bea
2 \cosh u &=& \mathrm{Tr} \left( e^A \right) = 2 \cosh \a , \nonumber \\
2 \cosh v &=& \mathrm{Tr} \left( e^B \right) = 2 \sinh \left( \lam /2 \right) \left[ \cosh \b \sinh \left( \r / 2 \right) + \a_1 \sinh \b \cosh \left( \r / 2 \right) \right] ,
\eea{inter_coord}
we can express $\b$ in terms of $v$, $\lam$, $\a_1$, $\a_3$ and $\r$.
The form of the exterior derivative of $\b$ is
\beq
d \b = \mathcal{C}_v d v + \mathcal{C}_\a d \a ,
\eeq{db}
because $\a_1$ and $\a_3$ are functions of $\a$, and $\r$ is not a modulus.
From the expression of the coordinates~(\ref{inter_coord}) it is obvious that $d \a = d u$, so 
only the coefficient of $d v$ is needed.
By expressing $\mathcal{C}_v$ in terms of $x$, $y$ and $z$ and using constraint relation (\ref{region}), we finally have
\beq
\w = \frac{k}{2\p} \frac{ dx \wedge dy }{xy - 2z}.
\eeq{symp_xyz}
This is the Weil-Petersson symplectic form given in~\cite{wol83}, up to a constant factor.
The symplectic form (\ref{symp_xyz}) was originally obtained in~\cite{wol83} for a torus with a puncture, 
but it is also valid for a torus with a hole.
By definition the Fricke-Klein coordinates obey $x,y,z>2$, so expression~(\ref{symp_xyz}) for the Weil-Petersson form is well-defined over all \tch\ space.

\section{K\"ahler Quantization on $T(\S)$}
\begin{figure}[h]
\begin{center}
\epsfig{file=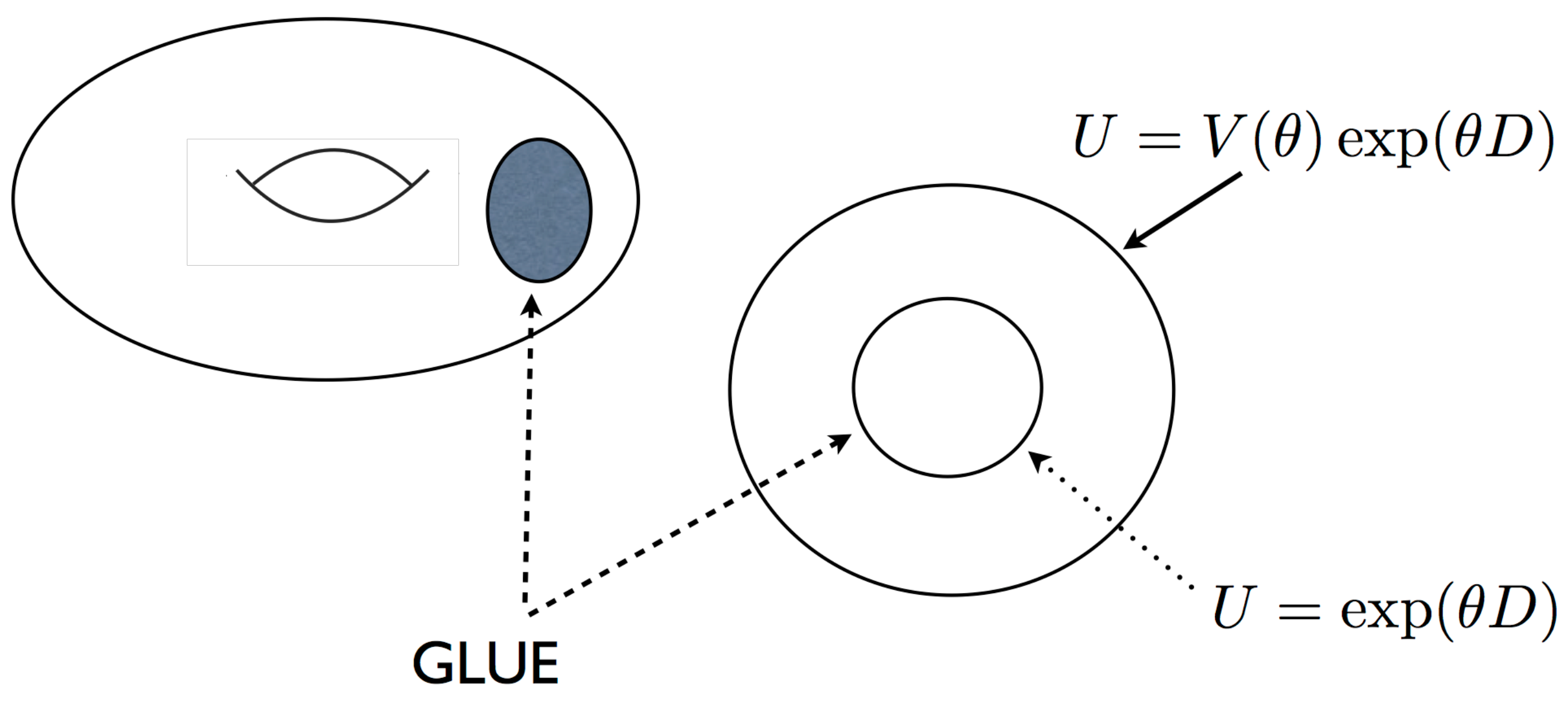, height=2in, width=4.5in}
\end{center}
\caption{Gluing a collar to $\S$.}
\label{collar}
\end{figure}

The parametrization $\tilde{A} = g^{-1} \td{d} g + g^{-1} H g$ , where $H$ is a flat $\slr$ connection obeying 
$H|_{\partial \S} =Dd\theta$, where $0\leq \theta <2\pi$ is the boundary coordinate, 
$D$ is a constant hyperbolic element of the $\widehat{SL}(2,\mathbb{R})$  algebra and $g$ is a single valued function 
from the Riemann surface $\S$ to the group $G$, reduces the CS 
action to eq.~(\ref{action_eff}). Geometrically, the three terms in the action come from decomposing the surface $\S$ into 
a ``collar'' region and a bulk. The collar is 
an annulus with one boundary coincident with the boundary of $\S$ and the other glued to the bulk as in 
figure~(\ref{collar}). Along the gluing cycle the gauge field is equal to $T^{-1} dT$ while on the other boundary it equals
$T^{-1}V^{-1} dVT$, $T=\exp(\theta D), V(\theta)\in \widehat{SL}(2,\mathbb{R})$. 

The first line in~(\ref{action_eff}) is
\beq
 I_1=\frac{k}{4 \p}\int_{R}{dt \int_{\pd \Sigma}{ \mathrm{Tr} \left( g^{-1} \td{d} g g^{-1} \dot{g} \right) }} + \frac{k}{12 \p}\int_{M}{\mathrm{Tr} \left( g^{-1} d g \right)^3}.
 \eeq{m11}
 This is the action of a chiral WZW model. By itself,  it would be invariant under the global symmetry 
 $g\rightarrow \omega(\theta)g$  and under the gauge symmetry $g\rightarrow g \omega(t)$~\cite{MS,p&s}. The first property 
 says that the Hilbert space of the quantized theory falls into representations of the Ka\v{c}-Moody algebra $\widehat{SL}(2,\mathbb{R})$, 
 while
 the second says that only representations containing a vector invariant under $\slr$ appear in the Hilbert space.
 In other words, the Hilbert space is made of the Ka\v{c}-Moody vacuum and its current algebra descendants. 
 In~(\ref{action_eff}), the term
 \beq
 I_2=-\frac{k}{2 \p}\int_{R}{dt{\Tr\int d\theta \left[ g^{-1} \dot{g} D(t) \right]}}
\eeq{m12}
breaks the gauge symmetry. The reparametrization $g\rightarrow g\exp \vartheta(t) $ with $\vartheta$ in the commutant of $D$
shows that $D$ and $\vartheta$ are canonically conjugate variables~\cite{MS}. For $\slr$ the commutant is one-dimensional.
One can pick $D$ as coordinate and $\vartheta$ as momentum. $D$ is either a nilpotent matrix or a diagonalizable one. In the latter case  wave functions are square summable functions of the
eigenvalue $\lambda$ of $D$. It is either a positive real number $\lambda \in \mathbb{R}$ or a root of unity.  
The Hilbert space arising from quantization of $I_1+I_2$ then contains Ka\v{c}-Moody representations
labeled by $\lambda$.  
By its definition $\lambda$ is also the charge along $D$ of the lowest-weight vector 
in the representation.  

 To have an Affine-Lie algebra $\widehat{SL}(2,\mathbb{R})$ one must impose chiral boundary conditions 
 $A_t- A_\theta=0$  (or $A_t+A_\theta=0$). Then the equations of motion imply that the nonzero boundary field is a
 function only of either $w^+=t/l+\theta$ or $w^-=t/l-\theta$. On the other hand, to find an asymptotically AdS$_3$ space
 obeying Brown-Henneaux~\cite{BH} boundary conditions, the $\slr$ fields must be further constrained~\cite{chvd}.
 The Brown-Henneaux boundary conditions are
\beq
g_{tt}= -r^2/l^2 + O(1), \qquad g_{t\theta}=O(1) , \qquad g_{tr}=O(r^{-3}), 
\eeq{m12a}
\beq
g_{rr}=l^2/r^2 + O(r^{-4}), \qquad g_{r\theta}=O(r^{-3}), \qquad
g_{\theta\theta}=r^2 + O(1).
\eeq{m12b}
These boundary conditions are preserved by diffeomorphisms with asymptotic 
form 
\bea
\zeta^t&=&l[f(w^+) + g(w^-)] + {l^3\over 2r^2} [\partial^2_+ f^+(w^+) + 
\partial^2_-g(w^-)] + O(r^{-4}), \nonumber \\
\zeta^\theta&=&[f(w^+) - g(w^-)] - {l^2\over 2r^2} [\partial^2_+ f^+(w^+) - 
\partial^2_-g(w^-)] + O(r^{-4}), \nonumber \\
\zeta^r&=& -r[\partial_+f(w^+) + \partial_-g(w^-)] + O(r^{-1}).
\eea{m12c}
The allowed diffeomorphisms are parametrized by two arbitrary functions $f(w^+)$, $g(w^-)$, 
each depending on only one of the 
two boundary light-cone coordinates; the boundary is at $r=\infty$. 

 With an appropriate gauge choice, the nonzero $\slr$ gauge field $A^+$ 
 preserving boundary condtitions~(\ref{m12a},\ref{m12b}) must have the form
 \beq
 A^+_+dw^+ = \begin{pmatrix} 0 & L(w^+)  \cr 1 & 0 \end{pmatrix} ,
 \eeq{m13}
 while the other $\slr$ gauge field, $A^-$, obeys a similar condition
 \beq
 A^-_- dw^-= \begin{pmatrix} 0 & L(w^-)  \cr 1 & 0 \end{pmatrix}.
 \eeq{m14} 
 Constraints~(\ref{m13},\ref{m14}) reduce the global symmetry of the combined action $I_1+I_2$ to a Virasoro algebra with
 (classical) central charge $c=3l/2G$~\cite{chvd}, set $\lambda \in \mathbb{R}$ and relates $\lambda$ to the 
 (semiclassical) weight $\Delta$ of the ground state in the Virasoro representation. The relation between asymptotic
 dreibein $e$, connection $\omega$ and $\slr$ gauge fields gives~\cite{bbo}
 \beq
 \Tr P e^{\int_{\partial \S} A} = 2 \cosh 2\pi\lambda = 2\cosh \left(2\pi \sqrt{ 3\Delta /c -1/8}\right).
 \eeq{m14a}
 
 This reduction to Virasoro symmetry can be seen either after~\cite{chvd}
 or before combining the two $\slr$ sectors (see appendix A of~\cite{hms}). 
 
 Ref.~\cite{hms} also points out to a fact 
 that has often been overlooked in the literature on 3D gravity. This is the role of Liouville theory in a conjectured
 AdS/CFT duality for pure 3D gravity~\cite{car}. If we {\em assume} that the 3D space is
topologically global AdS$_3$ we can reduce the CS action on the constraints, follow the Hamiltonian reduction
procedure of~\cite{chvd} and find a pure Liouville action. But in the presence of black holes, i.e. horizons, or of time-like
singularities associated with point-like particles in the bulk, the action at the $r=\infty$ boundary 
must be supplemented with 
other terms at the inner boundary/horizon. This was explicitly noted in~\cite{hms}. A possible interpretation of these terms 
is that they describe the states of the AdS$_3$ quantum gravity; more precisely the primary states in each irreducible 
representation (irrep) of the Virasoro$\times$Virasoro algebra acting on
the Hilbert space of quantum AdS$_3$ gravity\footnote{Ref.~\cite{MS} reviews
the ``constrain first'' Hamiltonian formalism to study  the effect
of point-like insertions and nontrivial topology for compact-group Chern-Simons theories.}. 
The role of the boundary Liouville theory would be then only to describe, in 
each irrep, the Virasoro descendants (cfr.~\cite{mart}). In this interpretation, 
other information is needed to determine the spectrum of primary operators. 

In our construction, the additional information needed to determine the theory is contained in the third term in 
action~(\ref{action_eff})
\beq
I_3= - \frac{k}{4 \p} \int_{R}{dt\int_{\S}{\mathrm{Tr} \left( H \wedge \dot{H} \right)}} .
\eeq{m15}
Since the boundary value of the flat connection $H$ is fixed to $Dd\theta$, the moduli space of $H$ is finite dimensional.
Its quantization, with a symplectic form determined by action $I_3$ and a choice of complex structure that makes such
moduli space a K\"ahler manifold, will produce a Hilbert space, whose vectors will label the different Virasoro irreps 
appearing in the Hilbert space. 

\subsection{The Teichm\"uller Component of Moduli Space}

The  previous section showed explicitly how (a component of) the moduli space of flat $\slr$ connections on the torus with
a single boundary component coincides with its Teichm\"uller space. The same section also showed 
that the symplectic form induced by action $I_3$ 
is the Weil-Petersson symplectic form. This is true in general because the $SU(1,1)$ K\"ahler form
\beq
\Omega =  \frac{k}{4\p}\int_{\S} d^2z {\left(  -\delta \w_z \wedge \delta \w_{\bar z}  + \delta e^+_z \wedge \delta e^-_{\bar z} \right) }, \qquad e^-_z=\overline{e^+_{\bar{z}}},
\eeq{m16}
reduces to the Weil-Petersson form after imposing the flatness condition $F=0$ and after  quotienting by gauge 
transformations. A proof of this fact  can be found in e.g.~\cite{hitch}. 

The relation to the metric formulation requires to define two $SU(1,1)$ connections in terms of metric data as~\cite{ks}
\beq
A_z^\pm=\frac{1}{2}\begin{pmatrix} \partial \phi &   \mp e^{\phi} \cr ie^{-\phi} q & -\partial\phi \end{pmatrix}, \qquad
A_{\bar{z}}^\pm=\frac{1}{2}\begin{pmatrix} -\bar\partial \phi &   -ie^{-\phi}\bar{q} \cr \mp e^{\phi}  & \bar\partial\phi \end{pmatrix},
\eeq{m17}
where $\phi$ and $q$ are defined as in eqs.~(\ref{m6},\ref{m7}). Up to a global gauge transformation both parametrizations
are  the same as that used [for a single $SU(1,1)$] in~\cite{V}. 
They are also the local form of a parametrization used by Hitchin~\cite{hitch} to describe
the Teichm\"uller slice of the moduli space of $\slc$ connections. In fact the parametrization used by Hitchin is the same as
in~\cite{V}. In local form in our case it reads
\beq
A_z^\pm=\frac{1}{2}\begin{pmatrix} \partial \varphi &   \mp e^{\bar\varphi} \cr ie^{-\bar\varphi} q & 
-\partial\varphi \end{pmatrix}, \qquad
A_{\bar{z}}^\pm=\frac{1}{2}\begin{pmatrix} -\bar\partial \bar\varphi &   -ie^{-\varphi}\bar{q} \cr \mp e^{\varphi}  & 
\bar\partial\bar\varphi \end{pmatrix}.
\eeq{m18}
The complex field $\exp\varphi$ is the local form of the canonical bundle $K$ of $\S$. To compute the symplectic structure
 it is convenient to choose chart coordinates such that $q=0$ and $\phi=\Re \varphi $ obeys a Liouville equation. The tangent vector
$\dot A^\pm$ decomposes into the sum of two terms, one proportional to the change in $q$, $\dot q$, 
the other proportional
to a non-holomorphic change of coordinates $z\rightarrow z'=z+ w(z,\bar{z})$. The latter obeys the infinitesimal form of 
the Laplace-Beltrami equation, $\bar\partial w= \mu(\bar{z})\exp(-2\phi)$, therefore $\mu$ is an antiholomorphic 
quadratic differential. The transformation 
$\varphi \rightarrow \varphi'= \varphi -(w\partial + \bar w \bar \partial) \varphi -\partial w$  {\em almost} cancels the change in 
$A^\pm$ due to the coordinate transformation, leading to the expression
\beq
\dot A^\pm  = \begin{pmatrix} 0 & e^{-\varphi} dz (\mp \bar{\mu} -i \dot {\bar q}) \cr 
e^{-\bar\varphi} d\bar z (\mp {\mu} +i \dot  q) & 0 \end{pmatrix} . 
\eeq{m19}
So $\dot A^\pm$ depends only on the quadratic differential $a^\pm=(\mp {\mu} +i \dot  q)$. 
The CS symplectic form~(\ref{m16}) reduces then to  the term
\beq
{k\over 4\pi}\int d^2 z e^{-\varphi -\bar\varphi} |a^\pm |^2,
\eeq{m20}
which is the Weil-Petersson symplectic form for the Teichm\"uller space defined by $\delta a^\pm$. An alternative proof 
of the equivalence between CS and Weil-Petersson forms can be found in~\cite{ks}. There, the equivalence is shown for 
the universal Teichm\"uller space, that contains the \tch\ spaces of any surface $\S$ as subspaces, defined by requiring that
the quadratic differentials transform appropriately under the $\slr$ transformations that define a Fuchsian 
uniformization of $\S$.
\subsection{K\"ahler Quantization}
Canonical quantization requires more than  a symplectic form, because a quantum theory requires a Hilbert space, that is
a scalar product. When the phase space admits a K\"ahler structure with K\"ahler potential $K$, the scalar product is 
given by the integral over phase space of $\exp(-K)\phi^* \psi$, as explained in the introduction after eq.~(\ref{constr}). The
wave functions $\phi,\psi$ are holomorphic section of a line bundle with curvature $\partial \bar\partial K$. Teichm\"uller 
space is noncompact so there is no quantization condition on the curvature. The Weil-Petersson form is not only 
symplectic but also K\"ahler. Its potential can be described quite explicitly in terms of a Liouville action~\cite{zo-tak}. 
In a given local chart $U_a$ of the surface $\S$ the 
Liouville field is defined as the scale factor of the metric  
\beq
ds^2_a = e^{2\phi_a} dz_a d\bar{z}_a .
\eeq{m21}
By construction it does {\em not} transform as a scalar. Instead, $\exp(2\phi)$ transforms as a density. 
So on the intersection of two charts, $U_a$ and $U_b$, the Liouville field obeys 
\beq
\phi_a = \frac{1}{2} \log |\partial z^b/\partial z_a|^2 + \phi_b.
\eeq{m22}

To define the action we need either to write it separately in each chart or  to specify a uniformization procedure for the 
surface. Ref.~\cite{zo-tak} employs the latter procedure and defines the action in terms of a Schottky uniformization. So $\S$ is
defined by cutting $2g$ disks $C_i$ out of the complex plane (or out of a disk ${\cal D}$  if a boundary is present) 
and by identifying the disks 
pairwise (say $C_i$ with $C_{g+i}$) by the $\slc$ maps $L_i$, $i=1,..,g$ defined by
\beq
{L_i(z) -a_i \over L_i(z) -b_i} = \lambda_i {z_i-a_i \over z_i - b_i} , \qquad 0< |\lambda_i| <1.
\eeq{m23}
Each $L_i$ maps the outside of one disk into the inside of the other; the region $D$ 
outside all disks (and inside ${\cal D}$ if a boundary is present) is $\S$. The Zograf-Takhtajan-Liouville 
action is then~\cite{zo-tak}
\beq
S_{ZTL}=\int_D |dz \wedge d\bar z| \left ( 2|\partial\phi |^2 + {1\over 2} e^{2\phi} \right) +i \oint_{C_i} \phi \left( dz {L''_i \over L_i'}-d\bar z {\bar{L}''_i \over \bar{L}_i'} \right) + \mbox{$\phi$-independent terms}.
\eeq{m24}
When evaluated on-shell, $-S_{ZTL}$ is the K\"ahler potential corresponding to the Weil-Petersson metric. While 
the Weil-Petersson metric is intrinsic to $\S$, the potential is not: it depends on the uniformization procedure. So one could define
other K\"ahler potentials that differ from~(\ref{m24}) by K\"ahler transformations: $K'= K + F + \bar{F}$, with $F$ a holomorphic function on \tch\ space. Examples of such potentials can be found in~\cite{tak-teo,yin}. The 
Schottky-uniformization potential has a nice geometrical interpretation in the case of closed Riemann surfaces: it computes
the regularized 3D volume of the handlebody $M$ with hyperbolic Einstein metric and constant-curvature 
boundary $\partial M= \S$~\cite{k}. We shall make use of this fact in section 6.

Eq.~(\ref{m24}) can contain parabolic singularities only. The extension to hyperbolic singularities is done by 
extending the surface with hyperbolic singularities (that is holes) as follows\footnote{This method is used e.g. in~\cite{wol-mirz} to 
prove the decomposition of the Weil-Petersson form given in eq.~(\ref{m42}) of section 5.1.}:  to each hole, a sphere with one hole 
and two parabolic singularities is
attached. The resulting surface contains only parabolic singularities (twice the number of hyperbolic ones). Its moduli space and 
symplectic form reduce
to \tch\ space and to the Weil-Petersson form by symplectic reduction~\cite{wol-mirz}. 

Eq.~(\ref{m24}) can also be defined for different uniformizations of the surface~\cite{tak-teo}. Finally, for the torus with one puncture 
explicit formulas for the Liouville action exist~\cite{men}. 

In spite of all that is known about the Liouville action, it would be desirable to have a definition of the Hilbert space of canonical gravity 
that does not depend on additional data 
 involving auxiliary 3D surfaces. The most important advantage of an intrinsic definition is that we will be interested in finding the 
 action of the mapping class group
 of $\S$ on wave functions. The Weil-Petersson metric defines a complex structure invariant under mapping class 
 group transformations. In such complex structure, mapping class group 
 transformations are holomorphic (see e.g.~\cite{wolp}). This nontrivial result, which is not shared by other 
 complex structures in \tch\ space, 
 means that a global diffeomorphism transforms $K$ by K\"ahler transformations: $K\rightarrow K'=K+F+\bar{F}$. This
 is good, yet transformations of $K$ under global diffeomorphisms are only known implicitly. They 
 depend as $K$ itself does on auxiliary 3D data. The solution to this problem is standard. In fact it has been used already in 
 the context of (pre)quantization of $\slc$ Chern-Simons theory~\cite{dey}.
 
 The trick is to use the freedom to redefine the K\"ahler potential in such a way that the new potential, 
 $\exp( -K')=\exp (-K)|H|^{-2}$,
 is only a function of 2D data. The holomorphic wave functions $\Psi$ are correspondingly redefined as
 \beq
 \Psi \rightarrow \Phi=H\Psi.
 \eeq{m25}
 In the case of \tch\ space, care must be exerted in defining such functions also on the boundary of the space. The appropriate
 holomorphic function $H$ is known thanks to the works of Quillen~\cite{quill} and Zograf~(unpublished, see~\cite{tak-lap}).
The function $H$, nonzero in the interior of $T(\S)$, is defined using the factorization formula for the determinant of the (nonzero eigenvalues of the) scalar Laplacian $\Delta$~\cite{tak-lap} 
\beq
{\det \Delta' \over \det \Im \Omega} = |G|^2 \exp (-S_{ZTL}),
\eeq{m26}
where $\Omega$ is the period matrix of the Abelian differentials on $\S$ and $G$ is holomorphic on \tch\ space 
(in fact on a quotient thereof, the Schottky space ${\mathfrak S}$~\cite{tak-lap}) and nonzero in its 
interior.\footnote{The decomposition in eq.~(\ref{m26}) depends of course on the definition of $S_{ZTL}$. 
In the following we will use an $S_{ZTL}$ that
vanishes at the boundary of \tch\ space. This definition differs from that of~\cite{tak-lap} by a K\"ahler transformation; the
corresponding function $G$ is holomorphic on \tch\ space but not on the Schottky moduli space.}

Eq.~(\ref{m26}) defines a metric on the space of holomorphic functions on $T(\S)$. It  transforms in the same way as 
$(\det \Im \Omega)^{-1}$ under mapping class group transformations. These are transformations on \tch\ space induced 
by large diffeomorphisms, that cannot be connected continuously to the identity. The explicit form of large diffeomorphisms
was mentioned in the introduction. They act by mapping a canonical basis of Abelian differentials, $a_1,..,a_g,b_1,..,b_g$
into another basis $a'_1,...,b'_g$. Denoting such canonical basis with $\underline{a},\underline{b}$, the transformation is 
\beq
\begin{pmatrix} \underline{a} \cr \underline{b} \end{pmatrix} \rightarrow \begin{pmatrix} \underline{a}' \cr \underline{b}' \end{pmatrix} 
= {\cal S}\begin{pmatrix} \underline{a} \cr \underline{b} \end{pmatrix}, \qquad
{\cal S}=\begin{pmatrix} A & B \cr C & D \end{pmatrix} .
\eeq{m27}
The intersection forms $\int a_i \wedge b_j=\delta_{ij}$, $\int a_i\wedge a_j =0$, $\int b_i \wedge b_j=0$ are topological 
invariants so they are preserved by all diffeomorphisms. This implies that the matrix 
$\cal S$ is real symplectic: $A^TC=C^TA$, $B^TD=D^TB$, $A^TD -C^TB=1$. The determinant of the Laplacian is a topological
invariant and $-S_{ZTL}$ is  the K\"ahler potential of the Weil-Petersson form with $c=2$ (see~\cite{tak-lap,zo-tak} for normalization conventions). $S_{ZTL}$ transforms under global diffeomorphisms as 
$S_{ZTL}\rightarrow S'_{ZTL}=S_{ZTL} - F_{c=2} - \bar{F}_{c=2}$ while $\det\Im \Omega$ transforms as 
$\det\Im \Omega \rightarrow \det \Im \Omega' = |\det(C\Omega + D)]|^{-2}\det \Im \Omega$.  The function $G$ thus transforms as 
\beq
G\rightarrow G'=\det(C\Omega + D) e^{-F_{c=2}} G .
\eeq{m28}

Since $G$ is holomorphic on $T(\S)$ and nonzero in its interior, it can be used to define new holomorphic wave 
functions and a new scalar product. If the functions $\Psi^I$ transformed as 
$\Psi^I \rightarrow \Psi'^{I}=U^I_J\exp(F_c)\Psi^J$ under
large diffeomorphisms, then the functions
\beq
\Phi^I=H\Psi^I, \qquad H=G^{c/2} ,
\eeq{m29}
will transform as
\beq
\Phi^I\rightarrow \Phi'^I=U^I_J{\det}^{c/2}(C\Omega + D) \Phi^J.
\eeq{m30}
The matrix $U^I_J$ relates wave functions defined in two coordinate systems on $T(\S)$.  For now we need not specify 
which property $U^I_J$ enjoys, besides the obvious one of being invertible.  

On the space of functions $\Phi$ redefined as in~(\ref{m29}) the scalar product is
\beq
\langle \Phi, \Phi'\rangle = \int_{T(\S)}\left({\det \Delta' \over \det \Im \Omega}\right)^{-c/2} \bar{\Phi} \wedge * \Phi '.
\eeq{m31}
The integrand in this formula is the Quillen norm~\cite{quill}. 

The scalar product defined by eq.~(\ref{m31}) is similar to that defined in~\cite{V,VV} for $\slr$ CS 
theory in the ``quantize first" approach and given in eq.~(\ref{inner_ver}).  In the semiclassical limit the two scalar 
products coincide up to $O(1)$ terms in the $1/c$ expansion, 
because the action $S_L - {\cal K}$ of eq.~(\ref{inner_ver}) is the same as $S_{ZTL}$ in eq.~\ref{m24}
and in the large $c$ limit the functional integral on $\varphi$ in~(\ref{inner_ver}) computes the action on the 
largest-action solution of the classical equations of motion. 

The Weil-Petersson metric is not the only one that gives a K\"ahler structure to \tch\  space. Another one, that makes 
$T(\S)$ into a metrically complete space, has been defined in~\cite{hitch}. It is obtained by 
restricting to $SU(1,1)$ connections a metric defined on flat $\slc$ connections ${\cal A}= A+iB$, $A,B \in SU(2)$. Such metric is in fact 
hyperk\"ahler with K\"ahler potential 
defined by restricting to flat connections the ``upstairs" potential~\cite{hitch}
\beq
K_H= \mbox{constant} \times \int_{\S} d^2z \Tr (A_{\bar z}  A_{z}  + B_z B_{\bar z} ) .
\eeq{m32}
The complex structure choice $A_{\bar z}, B_z$ induces a complex structure on the moduli space of $\slc$ connections.
On the $SU(1,1)$ subspace (in our definition $A,B$ are {\em antihermitian})
\beq
A_{\bar z}={1\over 2}\begin{pmatrix} \omega_{\bar z} & 0 \cr 0 & -\omega_{\bar z} \end{pmatrix} , \qquad
B_z=-{i\over 2}\begin{pmatrix} 0 & e^+_z  \cr  e^-_z & 0 \end{pmatrix},
\eeq{m33}
the potential becomes
\beq
K_{SU(1,1)}=\mbox{constant} \times \int_{\S} d^2z (\omega_{\bar z}  \omega_{z}  + e^+_z e^-_{\bar z} ).
\eeq{m34}
The K\"ahler metric induced on $T(\S)$ by $K_{SU(1,1)}$ differs from the Weil-Petersson one and so does the induced complex structure. On the other hand, the K\"ahler {\em form} coincides with the Weil-Petersson form, because
the latter is given by eq.~(\ref{m16}) and with an obvious choice of proportionality factor, eq.~(\ref{m34}) defines the form 
\beq
\Omega_{SU(1,1)}= {k\over 4\pi} \int_{\S} d^2z  (\delta \omega_{\bar z}  \wedge \delta \omega_{z}  + 
\delta e^+_z \wedge \delta e^-_{\bar z} ),
\eeq{m35}
which is the same as~(\ref{m16}) because the tangent vectors anticommute.

The upshot of this story is that the new K\"ahler structure defined in~\cite{hitch} generate a K\"ahler form symplectically equivalent to
 the Weil-Petersson form. Because of general properties of quantization this means that the Hilbert spaces obtained by quantizing the 
two structures are unitarily equivalent; therefore, we can limit ourselves to study one of them. We will work
 with the Weil-Petersson structure because it is much better understood and studied; besides, it is that structure that will
 allow us to make contact with conformal field theories. 
\section{The Hilbert Space of Quantum Gravity}
Quantization induced by the symplectic form obtained from the action $I_1+I_2$, given in eqs.~(\ref{m11},\ref{m12}),
 produces a Hilbert space 
 \beq
 {\cal H}=\int_{\mathbb{R}^+} d\lambda V_\lambda ,
 \eeq{m35a}
 with $V_\lambda$ a Verma module with lowest-weight vector $L_0|\lambda\rangle = \Delta | \lambda \rangle$. In the
 semiclassical limit $c=3l/2G\gg 1$ eq.~(\ref{m14a}) gives $\Delta=c\lambda^2/3 + c/24 + O(1)$. Wave functions are 
 \beq
 \mathfrak{v}=\int_{\mathbb{R}^+} d\lambda \phi(\lambda) v_\lambda, \qquad \phi \in L^2(\mathbb{R}^+), 
 \qquad v_\lambda \in V_\lambda .
 \eeq{m35b}
 The scalar product between two vectors $\mathfrak{v}=\int_{\mathbb{R}^+} d\lambda \phi(\lambda) v_\lambda$,
 $\mathfrak{u}=\int_{\mathbb{R}^+} d\lambda \psi(\lambda) u_\lambda$  is  defined in terms of the scalar
 product $(u_\lambda | v_{\lambda'})=\rho(\lambda)\delta(\lambda-\lambda')$ of the representation $V_\lambda$ as
 \beq
 \langle \mathfrak{u} | \mathfrak{v}\rangle = 
 \int_{\mathbb{R}^+} d\lambda \rho(\lambda) \bar\psi(\lambda) \phi(\lambda)  .
 \eeq{m35c}
 
 The last component of the Hilbert space results from the quantization of action $I_3$ in~(\ref{m15}). Wave functions are
 holomorphic sections defined on the finite-dimensional moduli space $T(\S)$ of flat connections with fixed value on
 $\partial\S$. Their scalar product is defined in eq.~(\ref{m31}). 
Because the action of classical gravity can be written as the sum of two $\slr$ CS theories and quantization of each one
gives a Hilbert space, we might think that the 
Hilbert space of quantum gravity is the product of the two.  This is not correct because of global diffeomorphisms, which we study next. 

\begin{figure}[h]
\begin{center}
\epsfig{file=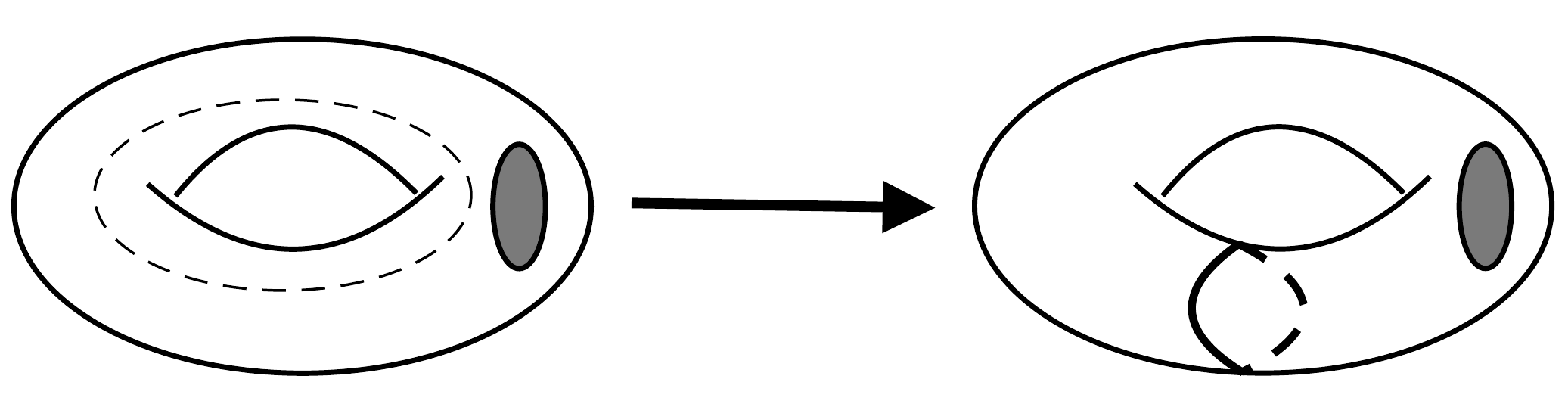, height=1in, width=4in}
\end{center}
\caption{A global diffeomorphism changes the homology basis of $\S$. Here the  cycle represented with a thin line 
transforms into that represented by a thick line. The change of basis also induces a change in the pant decomposition of 
the surface.}
\label{mcg}
\end{figure}

The canonical formalism restricts the topology of spacetime to $M=\S\times \mathbb{R}$. Global diffeomorphisms on 
$M$ preserve the topology $\S \times \mathbb{R}$ so they reduce to  one-parameter smooth families of 2D 
diffeomorphisms labeled by the time coordinate $t\in \mathbb{R}$. Similarly, when $M=\S \times I$, with 
$I\subset \mathbb{R}=[t_1,t_2]$ a time interval, the diffeomorphism acting on the initial surface $\S\times \{t_1\}$ is
continuously connected to the diffeomorphism acting on the final surface $\S\times \{t_2\}$. The description of phase
space of 3D gravity as a product of an ``initial" and a ``final" phase space, given in section 2, then suggests that global 
diffeomorphisms act diagonally on the two phase spaces. This is more than a suggestion because the same conclusion
can be reached also in the Chern-Simons formalism. In CS, the moduli space of flat solutions is parametrized by the 
holonomies (or Wilson loops) of two independent gauge fields, $A^\pm$. A global diffeomorphism $\phi: M\rightarrow M$ 
maps a cycle $\gamma$ into its image $\phi(\gamma)$ as shown in figure~(\ref{mcg}). 
Since there is a single basis of cycles for both gauge fields, both 
Wilson loops transform in the same manner 
\beq
\Tr P e^{\int_\gamma A^\pm} \rightarrow \Tr P e^{\int_\gamma \phi_*A^\pm} = \Tr P e^{\int_{\phi(\gamma)} A^\pm} .
\eeq{m36}
This equation shows that the mapping class group acts diagonally on the holonomies of the two factor gauge groups, 
so we conclude that the phase space of gravity in either the CS or metric formulation is:
\beq
{T(\S)\times T(\S)\over M}.
\eeq{m37}
Implementation of general covariance in a quantum theory of gravity requires that wave functions
be invariant under all diffeomorphisms, including those not connected to the identity. Those transformations are
mapping class group elements for manifolds $M=\S \times \mathbb{R}$. 
Global diffeomorphisms change the holonomy basis of $\S$; such transformations leave the Weil-Petersson symplectic 
form invariant and are thus canonical transformations which are realized by unitary transformations in the quantum theory.
The scalar product is defined by eq.~(\ref{m31}) so unitarity 
means that the matrix $U^I_J$ defined earlier obeys $\sum_L U^I_L \overline{U^J_L} =\delta^I_J$. 

With a bit of hindsight it is convenient to define the wave function on $T(\S)\times T(\S)$ as  holomorphic in the coordinates
$\mu$ of the first factor and antiholomorphic in the coordinates $\mu'$ of the second factor. 

The measure of integration for $\slr\times \slr$ is the product of the measures in eq.~(\ref{m31}), so it is not invariant 
under the diagonal mapping class group; instead, it transforms  as
\bea
&&{\det \Delta' \over \det \Im \Omega}(\mu,\bar\mu) {\det \Delta' \over \det \Im \Omega}(\mu',\bar\mu')
\rightarrow \nonumber \\ 
&&\rightarrow \left |{\det}(C\Omega +D)\right|^2 {\det \Delta' \over \det \Im \Omega}(\mu,\bar\mu) 
\left|{\det}(C\Omega +D)\right |^2 {\det \Delta' \over \det \Im \Omega}(\mu',\bar\mu').
\eea{m38}
So, invariance under mapping class group transformations means that the wave function transforms as the $c/2$ power 
 of the determinant line bundle ${\cal L}$ on $T(\S)$ times the $c/2$ power of the line bundle $\overline{\cal L'}$ on  $T'(\S)$. 
 On a point $\mu\in T(\S)$, the determinant 
bundle is the top exterior power of the space $H^0(\S,K)$ of holomorphic differentials. 

Concretely, because of transformation property~(\ref{m30}), on each \tch\ space the wave function is an entangled state of the form
\beq
 \Xi=\sum_{IJ} N_{IJ}\Phi^I (\mu ) \Psi^J (\bar\mu') .
\eeq{m39}
The coefficients $N_{IJ}$ are independent of $\mu,\mu'$ and must obey
\beq
\sum_{LM}U^L_I \overline{U^M_J}N_{LM} =N_{IJ},
\eeq{m40}
 so that $\Xi \in {\cal L}^{c/2}\times \overline{{\cal L'}^{c/2}}$. 
 If $\S$ had no boundary, then a section of  ${\cal L}^{c/2}\times \overline{{\cal L'}^{c/2}}$  would be naturally interpreted as
 a partition function of a CFT~\cite{w8,seg}. In the presence of a boundary a CFT interpretation is also 
 possible.
 
Once the dependence on the boundary degrees of freedom is made explicit, quantization of $\slr\times \slr$ produces
vectors
\beq
 \mathfrak{V}_g= \sum_{n\in \mathbb{Z}} \int_{\mathbb{R}^+}d\lambda 
 \int_{\mathbb{R}^+} d\lambda' \delta[\Delta(\lambda)-\Delta(\lambda') -n]\phi(\lambda,\lambda') 
 v_\lambda \otimes u_{\lambda'}  \sum_{IJ} N_{IJ}\Phi^I (\mu,\lambda ) \Psi^J (\bar\mu',\lambda').
 \eeq{m41}
 The restriction to $\Delta(\lambda)-\Delta(\lambda')$ integer is itself a consequence of diffeomorphisms invariance; the 
 diffeomorphism in question  is in this case a $2\pi$ rotation of the boundary coordinate. 
 
  \subsection{A Suggestive Form of the K\"ahler Potential}
 We can be more explicit about the contribution of the boundary of the surface $\S$ to the integration measure 
 in~(\ref{m31}), thanks to a theorem due to M. Mirzahani~\cite{mirz}. In our case it states the following: Let $\Omega_L$ be
  the Weil-Petersson form of a Riemann $\S$ surface with a boundary $\partial \S$ of geodesic length $L$  (in the metric of constant curvature $-1$). Let $\Omega_0$ be the form for a once-punctured surface; let the puncture be at $z\in \S$. Let 
  $\omega(z)$ be a representative of the first Chern class of the tautological line bundle over the \tch\ space of once-punctured surfaces. Then in cohomology 
 \beq
 [\Omega_L]= [\Omega_{0}] + {L^2\over 2} [\omega(z)].
 \eeq{m42}
 A proof of this equality, which employs the Duistermaat-Heckman theorem, can be found in~\cite{mirz} and in the lecture 
 notes~\cite{wol-mirz}. The \tch\ space $T_{g,1}$ of a once-punctured Riemann surface of genus $g$ is a bundle over the 
 \tch\ space  $T_g$ of surfaces $\S_0$ of the same genus but without punctures. The fiber at $m\in T_g$ is $\S_0(m)$. 
 The tautological line bundle over $T_{g,1}$ is the tangent space of $\S_0(m)$ at $z$. A representative of $[\omega(z)]$ is 
 then simply the curvature form $R$  of $\S_0$ at a point $z\in \S_0(m)$ (see e.g. eq. [5.40] in~\cite{dvv}). 
 In complex coordinates 
 the 2D metric at $z$ is $ds^2= \exp[2\phi (z)]dz d\bar z$ so $R=-\partial\bar\partial \phi (z) $. This choice identifies the 
 K\"ahler potential of surfaces with geodesic boundary of length $L$ as
 \beq
 K_L(m,\bar m, z,\bar z)= K_0(m,\bar m)  -  k L^2 \phi(z) - F(m) - \overline{F(m)}.
 \eeq{m43}
 The holomorphic function $F$ can be reabsorbed in a wave function redefinition\footnote{Of course the old wave 
 function $\Psi$ will transform differently, under a change of homology basis, than the new one, $\Phi=\exp(F) \Psi$.}. 
 Equation~(\ref{m43}) now introduces an explicit dependence of the measure of integration on the 2D metric. It also
 renders explicit the dependence on the geodesic length. 
 
 \subsection{CFT Interpretation}
 We are at last ready to give a conformal field theory interpretation to wave function~(\ref{m41}). It is written as an
 integral of vectors living in representations of Virasoro$\times$Virasoro, which are labeled by $(\lambda,\lambda')$. 
 The integrand, $\Xi$ in eq.~(\ref{m39}) transforms under the diagonal action of the mapping class group as an element of 
 ${\cal L}^{c/2}\times \overline{{\cal L'}^{c/2}}$, that is as in eq.~(\ref{m3}).
 
 Thanks to eq.~(\ref{m43}) we can define a new object
 \beq
 \Xi_{L,L'}' = \exp[ F(m) + \overline F(m')] \Xi,
 \eeq{m44}
 which now transforms under $M$ in the following manner.
 
 On the surface $\S$, a mapping class group transformation that leaves the point $z$ invariant transforms the metric at $z$
 as $ds^2\rightarrow |\Omega(z)|^2 ds^2$, for  some nonzero holomorphic function $\Omega$. The factor 
 $\exp [kL^2 \phi(z)]$ in $\exp (-K_{L})$ makes the measure transform as 
 $\exp (- K_{L})\rightarrow |\Omega|^{kL^2} \exp (- K_{L})$. The same happens with the other term in the measure, 
 $\exp (-K_{L'})$. 
 The function $\Xi$ transformed as in~(\ref{m3}) so the new function $\Xi'$ now transforms as
 \beq
 \Xi' \rightarrow \Omega^{kL^2/2}(z) \bar\Omega^{kL'^2/2}(\bar z) {\det}^{c/2}(C\Omega(\mu) + D) 
 {\det}^{c/2}(C\overline{\Omega}(\bar{\mu}') + D)\Xi'.
 \eeq{m45}
In a CFT, this is the transformation law for the vacuum expectation value on $\S_0$ of a primary operator $O(z,\bar z)$ with 
conformal weights $\Delta=kL^2/2$, $\Delta'=kL'^2/2$. Confronting these formulas with the expressions given above 
eq.~(\ref{m35b}), $\Delta=c\lambda^2/3 +c/24 + O(1)$ etc., we identify $kL^2/2=c\lambda^2/3 + O(1)$. The $L$-independent 
shift in the conformal dimension can be accounted for by an $L$-independent redefinition of the wave function.

\subsection{Normalizability Constraints}
\begin{figure}[h]
\begin{center}
\epsfig{file=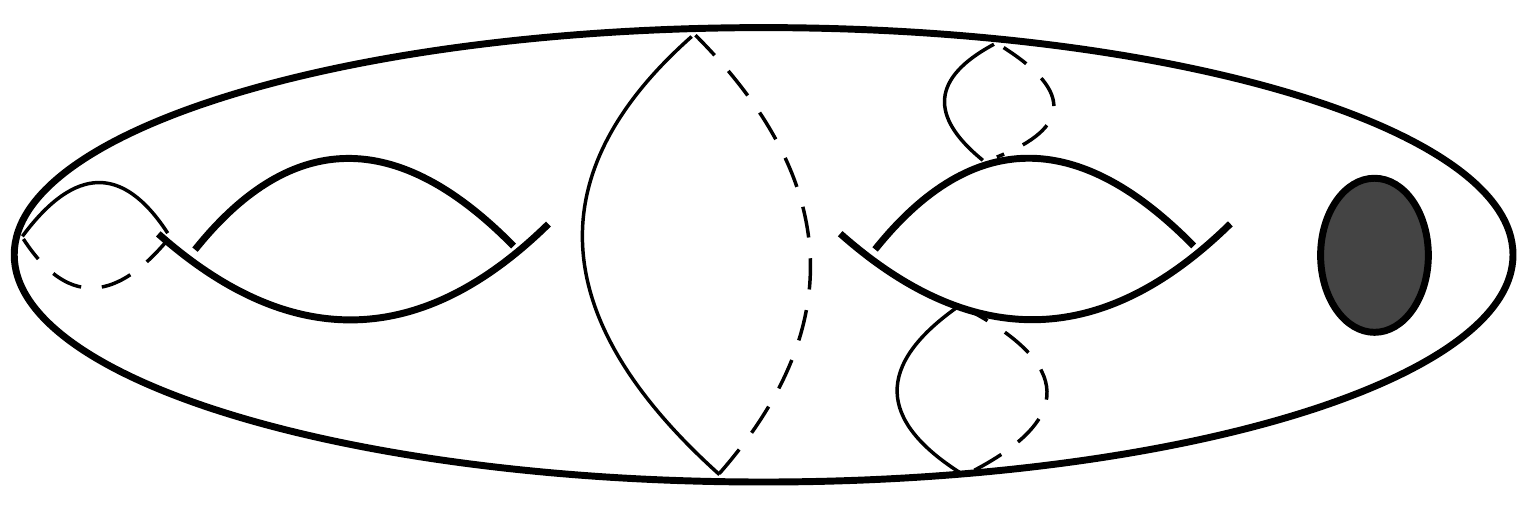, height=1in, width=3.5in}
\end{center}
\caption{Pant decomposition of  a $g=2$, one-boundary Riemann surface}
\label{pant}
\end{figure}

Normalizability of the wave function $\mathfrak{V}$ imposes certain constraints on $\Xi$ that translate into 
constraints on its CFT interpretation. Specifically, from the condition 
$|| \mathfrak{V} ||<\infty$ it follows that $\Xi$ is plane-wave normalizable, which in turns means
\beq
\int_{{T(\S)\times T(\S)\over M}} \left({\det \Delta' \over \det \Im \Omega}\right)^{-c/2} 
\left({\det \Delta' \over \det \Im \Omega}\right)^{-c/2}\overline{\Xi} \wedge * \Xi \equiv || \Xi ||^2 < \infty , \qquad
\Xi=\sum_{KL} N_{KL}\Phi^K (\mu ) \Psi^L (\bar\mu').
\eeq{m46}
These normalizability conditions become more intelligible when $[T(\S)\times T(\S)] / M$ is parametrized in terms of
Fenchel-Nielsen (FN) coordinates~\cite{wolp,wol-WP} relative to a pant decomposition of $\S$. The example in figure~(\ref{pant}) is a one-boundary surface of genus $g=2$ obtained by gluing together three pants (aka trinions) along their boundaries. 
Each trinion is topologically a sphere with 3 holes. The geodesic lengths $l$ of 
each hole --in a metric of constant curvature $R=-1$--  make up  half of the  the FN coordinates. The other coordinates can 
be normalized in such a way to become the angle by which two holes are twisted relative to each other, 
before being glued together. In this normalization the other sets of coordinates are thus angles $\theta$. 
The WP form is then~\cite{wolp,wol-WP}
\beq
\Omega_{WP}= \sum_i{l_i\over 2\pi} dl_i \wedge d\theta_i, \qquad i=1,...., 3g-3+1 .
\eeq{m47}
While the WP form is simple, the complex structure defined by the WP {\em metric} is not. Near a degenerate 
surface, that is a boundary of moduli space, asymptotic formulas for the WP metric are known~\cite{wolp,mazz}. For us
it is sufficient to know that near a degeneracy point, where one or more of the lengths $l_i$ go to zero, the correct complex 
coordinate is
\beq
q_i = \exp (2\pi i\tau_i)=\exp(-2\pi^2/l_i +i\theta_i), \qquad \tau_i=i\pi/l_i +\theta_i/2\pi,
\eeq{m48}
so that the WP form becomes 
\beq
\Omega_{WP} \approx {4\pi^2\over (\tau_i -\bar{\tau}_i)^3}  d\tau_i \wedge d\bar{\tau}_i .
\eeq{m49}
The pinch is at $\Im \tau_i =\infty$ and there the K\"ahler potential for 
$\Omega_{WP}$ is $K= -2\pi^2/(\tau_i -\bar\tau_i) + 2\Re F+... $ where the ellipsis stand for terms decaying faster 
than $\Im \tau^{-1}$ and $F$ is a holomorphic 
function of $\tau_i$. So, at the pinch, the normalizability of the wave function  depends crucially on this unknown $F$, 
which cannot be determined by integration of the K\"ahler form. 

\begin{figure}[h]
\begin{center}
\epsfig{file=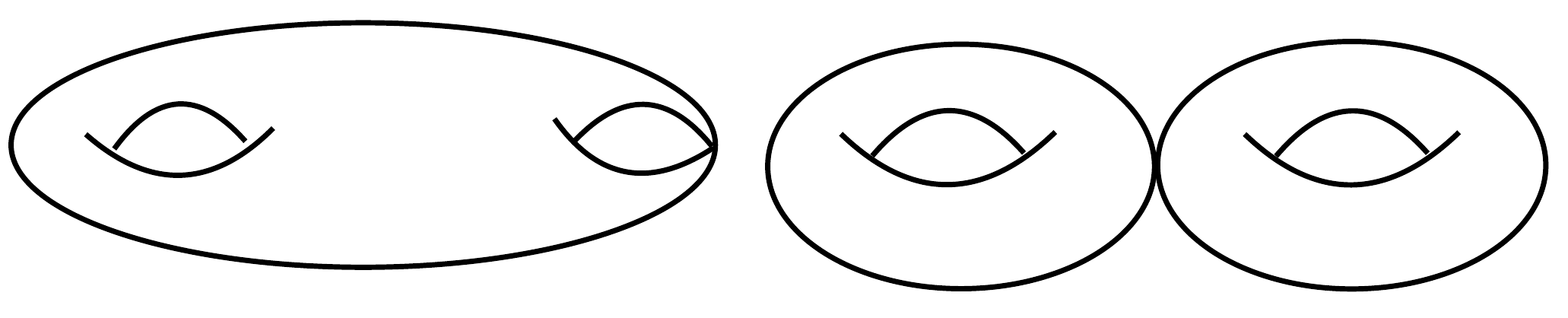, height=1in, width=5in}
\end{center}
\caption{Pinching  a $g=2$ Riemann surface. On the left, a non-separating pinch; on the right, a separating pinch}
\label{pinch}
\end{figure}

Luckily, we have a better formula for the integration measure in eq.~(\ref{m46}), which is known precisely near the pinch
$q=0$ thanks to ref.~\cite{wol-lap} (see also~\cite{cmnp}). The result depends on whether the pinch is along a
separating geodesic as in the right image in figure~(\ref{pinch}), or along a 
non-separating geodesic~(see e.g.~\cite{iengo}), as in the left image. 
\beq
{\det \Delta' \over \det \Im \Omega}(q,\bar q,...) \rightarrow  C (q\bar q )^{1/12}\left\{ \begin{array}{ll} 
1 &\mbox{separating geodesic}\\ -\log |q| & \mbox{nonseparating geodesic} \end{array} \right. ,
\eeq{m50}
with $C$ a nonzero function of the other moduli. 
Using standard properties of the Laplace transform we can write the wave function $\Xi$ near a pinching geodesic as 
\beq
\Xi=\int d\Delta\int d\bar\Delta' \tilde{\phi}(\Delta,\bar\Delta') q^\Delta \bar{q}'^{\bar\Delta'} .
\eeq{m51}
The natural measure of integration in \tch\ space, $\Omega_{WP}^{3g-3+1}$, becomes proportional to powers of 
$\Im \tau$ and $\Im \tau'$  times
$d^2\tau d^2\tau'$ near the $q=0$ ($\Im\tau \rightarrow \infty$) boundary. To compute the contribution of the region 
$|q|\ll 1$ to the integral~(\ref{m46}) we need still to find the region of integration in $[T(\S)\times T(\S)]/M$. The Fenchel-Nielsen coordinate range that covers $T(\S)$ is $l\geq 0$, $-\infty < \theta < +\infty $. The mapping class group identifies
$\theta\sim \theta + 2\pi$. While all mapping class group elements can be written as combinations of $2\pi$ twists 
along {\em some} geodesics, their form for a given pant decomposition may be in general complicated. The element that
acts simply as a $2\pi$ twist along the degenerating geodesic will be enough to find new constraints on $\Xi$. 

The diagonal mapping class group transforms $(\theta ,\theta')\rightarrow (\theta+2\pi,\theta'+2\pi)$, where
$\tau=i\pi/l+\theta/2\pi,\tau'=i\pi/l'+\theta'/2\pi$, so the variable $\phi=(\theta+\theta')/2$ is periodic with period $2\pi$ while
the variable $\psi=(\theta-\theta')/2$ runs over all $\mathbb{R}$. Periodicity in $\phi$ constrains $\Xi$ to have the form
\beq
\Xi=\sum_{n\in \mathbb{Z}}\int d\Delta  \tilde{\phi}(\Delta,\Delta +n) q^\Delta \bar{q}'^{\Delta+n} .
\eeq{m52}

For large $\Im \tau, \Im\tau'$ (say both larger than a constant $K\gg 1$) the integral~(\ref{m46}) can be evaluated as follows. 
Using a result of L. Bers, ref.~\cite{wol-mirz} shows that there exists a constant $C<K$ such that the region $\Im \tau \geq C$,
 $0\leq \phi < 2\pi$ is a rough  fundamental domain of \tch\ space. In our case this property implies that there exists a finite integer $N$ such that,
under the action of the diagonal modular group $M$, each point in $[T(\S)\times T(\S)]/M$ has at most $N$ images in 
the domain. The domain of integration $\Im \tau >K$, $\Im \tau' >K$ is obviously a subset of the 
rough fundamental domain. Performing the integral~(\ref{m46}) on such domain will give a result smaller than 
$N || \Xi ||^2$. 
Then the $N$-multiple of the integral in eq.~(\ref{m46}) bounds from above another integral, that can be done explicitly; to wit:
\bea
&& \int d\Delta \int d\Delta' \sum_{n,m}\int_{-\infty}^{\infty} d\psi \int_0^{2\pi} d\phi \int_K^{\infty} d\Im\tau \int_K^{\infty} d\Im\tau'  F(\tau,\tau')
\tilde{\phi}(\Delta,\Delta +n)\overline{\tilde{\phi}(\Delta',\Delta' +m)}\times \nonumber \\
&& \times 
e^{i \psi (2\Delta -2\Delta' +n-m) +i\phi(m-n) -2\pi \tau_I (\Delta +\Delta' -c/12) -2\pi \tau_I' (\Delta +\Delta' +n +m-c/12)} < N ||\Xi ||^2 < \infty.
\eea{m53}
Here $F(\tau,\tau')$ is a real, positive polynomial function that takes into account contributions to the measure 
non-exponential in $\tau$, such as those in eq.~(\ref{m50}). 

We get two constraints from this equation. 

The first is obtained by integrating over $\psi$ and $\phi$. Integrating over $\phi$ sets $n=m$ and 
integrating over $\psi$ result is a delta function 
$\pi \delta(\Delta -\Delta')$. It equates $\Delta$ to $\Delta'$ and forbids the presence of a distributional 
component proportional to $\delta(\Delta -\mbox{constant})$ or $\Delta$-derivatives thereof in 
$\phi$. Such terms would make the norm of $\Xi$ infinite. They would also transform the integral in 
$\Delta,\Delta'$ into a sum. 
We can say then that the first constraint forbids the presence of a discrete component in the spectrum of $\Delta$. 

The second constraint follows from the fact that the exponentials in $\tau_I,\tau'_I$ must be non-negative, since for $n=m$ the integrand in $\Delta$ is a sum of manifestly positive terms. It is 
$\Delta \geq c/24$, $\Delta+n\geq c/24$. 

We thus arrive at the most important result of our analysis so far. In view of the interpretation of $\Xi$ [or better $\Xi'$ in
eq.~(\ref{m44})] as the VEV, on the surface $\S_0$, of a CFT primary operator, the integral in~(\ref{m51}) is the sum over 
conformal field theory states $|\Delta,\Delta'\rangle$ flowing across the degenerating geodesic (the pinch). The constraint 
$\Delta=\Delta'+n$,  $n\in \mathbb{Z}$ becomes $(L_0-\bar{L}_0)|\Delta,\Delta'\rangle= n |\Delta,\Delta'\rangle$ while
the constraints following from condition~(\ref{m53}) say that the states belong to a CFT with 
continuous spectrum and possessing a 
gap on the dimension of primary operators $\Delta\geq c/24$.\footnote{The domain of integration in eq.~(\ref{m46}) extends to the whole region $\Im \tau' \geq 0$. 
In principle one could get extra constraints on the wave function by studying the region near $\Im \tau'=0$, but we have not 
been able to find interesting constraints so far.}

This result must be qualified because $\Xi'$ may be a linear superposition of operators 
with the same conformal dimension, but belonging to different CFTs. Still, our result says that 
\begin{quotation}
The VEV $\langle O \rangle_\S $ of a CFT primary operator gives a physical (i.e. normalizable) vector in the Hilbert space
of pure $AdS_3$ gravity only if it belongs to a CFT with continuous spectrum and obeying a bound on primary 
operator dimensions $\Delta = c/24 + O(1)$. 
\end{quotation}
We added a possible $O(1)$ term in the bound because all our formulas for the measure of integration were derived for 
$c\gg 1$ and could receive corrections at finite $c$. A specific quantum-corrected form for the measure was 
proposed in~\cite{V,VV}. As we said earlier, it reduces to~(\ref{m31}) up to $O(1)$ terms in the $1/c$ expansion.  The explicit 
form of the measure of~\cite{V,VV} is, in standard notation
\beq
\langle \Phi,\Phi'\rangle_V= 
\int_{T(\S)}Z_L^{26-c}|\det {\partial_{K^{-1}} }|^2 \left({\det \Delta' \over \det \Im \Omega}\right)^{-c/2} \bar{\Phi}\wedge *\Phi'. 
\eeq{m54}
Here as in refs.~\cite{V,VV} the Liouville partition function $Z_L^{26-c}$ is defined in terms of a {\em scalar} 
Liouville field $\phi$ and a background metric $g^0_{\alpha\beta}$ with constant curvature $R=-1$ as 
$Z_L^{26-c}= \int [d\phi] \exp[(c-26)S_L(\phi,g^0)]$. 
The scalar $\phi$ transforms as $\phi\rightarrow \phi'=\phi-\log\Omega$ under a Weyl rescaling
$g^0_{\a\b}\rightarrow g_{\a\b}=\Omega^2 g^0_{\a\b}$ and the action $S_L$ can be chosen to vanish on $g^0_{\a\b}$. 
In the semiclassical limit the integrand prefactor that multiplies the Quillen norm reduces then to $q\bar{q}$ near a factorization
boundary. For Liouville theory on a torus, this property is due to the explicit form of the Liouville partition function on the
 torus~\cite{seib}, which is the same as that of a noncompact scalar. On a higher-genus surface the factorization 
 follows from the standard CFT relation between partition function and conformal blocks. 
 
 The contribution of the boundary is a bit hidden in the scalar product formula~(\ref{m54}), 
 but it can be made more explicit following the method that we used in subsection 5.1.  
 Liouville action can also be rewritten as a bulk action defined on a punctured Riemann surface $\S_0$,
 plus a contribution at the puncture $z_0$, in analogy with eq.~(\ref{m43}): $S_L=S_L^0 + \alpha\phi(z_0)$. When substituting in 
 eq.~(\ref{m54}) and recalling that ${\det}^{-1/2} \Delta'$ is the partition function of a free massless scalar, while 
 $|\det {\partial_{K^{-1}} }|^2$ is the partition function $Z_{bc}$ of  the bosonic-string $bc$ ghost system, 
 we get a suggestive form for the scalar product~(\ref{m54}):
 \beq
\langle \Phi,\Phi'\rangle_V= 
\int_{T(\S)}\langle e^{\alpha \phi (z_0)}\rangle_L^{26-c} (Z_S)^c  Z_{bc}(\det\Im \Omega)^{c/2}\bar{\Phi}\wedge *\Phi'.
\eeq{m54a}
 In the expectation value of the Liouville theory vertex $\exp \alpha\phi$,  $\alpha$  is only determined to leading order in $c$. Two equivalent choices are particularly interesting, namely 
 \beq
 \alpha_\pm= {Q\over 2} \pm  i\sqrt{ \Delta(\lambda)-{c-1\over 24}}, \qquad 1-6Q^2=26-c=c_L.
 \eeq{m54b}
 With either choice, the integrand in eq.~(\ref{m54a}) becomes identical to a 2D quantum gravity amplitude~\cite{dk,V,VV} because
 the dimension of the Liouville vertex, $\Delta_L=\alpha(\alpha -Q)$ and the dimension of $\Delta(\lambda)$ of $O$ add up to
 $\Delta_L+\Delta(\lambda)=1$\footnote{The 2D quantum gravity considered here is beyond the $c=1$ barrier~\cite{dk}, so 
 its definition is a bit formal because it requires an analytic continuation. The Liouville vertex too is only defined by analytic continuation. On the torus, explicit formulas for the Liouville vertex are known~\cite{had} and do reduce to the semiclassical ones for
 large $c$.}. 
 
 Using definition~(\ref{m54a}), the 
 bound on $\Delta$ also becomes a suggestive one, that we will adopt for the rest of the paper
\beq
\Delta \geq {c-1\over 24}.
\eeq{m55}

What we found so far can be summarized by saying that the Hilbert space of canonically quantized pure quantum gravity 
on the asymptotically $AdS_3$ space $\S \times \mathbb{R} $ is the direct product of irreducible representations of
Virasoro$\times$Virasoro times the target space of VEVs of primary vertices in {\em Liouville-like} conformal field theories. 
By Liouville-like we mean theories with a continuous spectrum of primary operators and no primaries with weight 
$\Delta<(c-1)/24$ (so in particular no $\slc$-invariant vacuum). 

When CS gravity is quantized on the space $\S \times \mathbb{R}$, with $\S$ a closed surface of genus $g>1$, it has been 
noticed by Witten~\cite{w8} that the resulting Hilbert space can be thought of as the target space for CFT partition functions.

Our result extends and qualifies Witten's in two ways. The first one is by extending the analysis to the case that $\S$ has a 
boundary. A boundary means that the Hilbert space is infinite-dimensional, but the infinite-dimensional part of the 
space factorizes as 
\beq
V_{\lambda,\lambda'}\otimes H_{\lambda,\lambda'},
\eeq{m56}
with $V_{\lambda,\lambda'}$ a unitary irrep of Virasoro$\times$Virasoro and $H_{\lambda,\lambda'}$ a finite-dimensional Hilbert space.
The second modification to Witten's result is that normalizability of the wave function restricts the class of CFTs admissible
in canonical quantization. 

To summarize, the Hilbert space one obtains from canonical quantization of pure gravity is 
\beq
\mathfrak{H}=\sum_{n\in \mathbb{Z}}\int_{\Delta, \Delta+n 
\geq (c-1)/24}d\Delta  \sum_{g=1}^\infty V_{\Delta,\Delta+n}\otimes H^g_{\Delta,\Delta+n} 
\oplus V_{\Delta=\Delta'=0}.
\eeq{m56a}
Here we more conveniently labeled the Virasoro$\times$Virasoro irreps by their lowest conformal weights $\Delta,\Delta'=\Delta+n$.
The sum over genera appears because the Hilbert spaces obtained by quantizing on surfaces of different genus 
are orthogonal in our construction. The $\slr\times\slr$-invariant vacuum and its Virasoro descendants  
belongs to the physical Hilbert space since they are obtained by quantizing the theory on $\S=$ a disk.

\subsection{Didn't We Know it Already?}
One last point that needs clarification is why we went to the trouble of studying in great details normalizability conditions 
and so on only to 
arrive at the result that the Hilbert space describes CFTs resembling Liouville theory. Hasn't it already been conjectured 
in~\cite{V,VV} and proven in~\cite{tesch,tesch2} that quantization of $\slr$ CS produces a Hilbert space whose wave 
functions are Virasoro conformal blocks that transform under modular transformations as Liouville conformal blocks? 
The point is that conformal blocks alone are not sufficient to determine a CFT. The same blocks can give rise to theories
with either continuous or discrete spectrum, for instance. The simplest example of this fact is given by the Virasoro 
conformal blocks at $c=1$. They can be combined into either the CFT of a noncompact free scalar, with continuous
spectrum, or a compact scalar with discrete spectrum. In either the algebraic approach or our own, one would
have to define states invariant under the diagonal mapping class group $M$ and redefine in an appropriate manner the
scalar product. In our K\"ahler ``coherent state" quantization, we implemented this redefinition by restricting
the domain of integration on $T(\S)\times T(\S)$ to the fundamental domain of $M$. So, in our approach to quantization, 
it is only when conformal blocks are combined 
together into the entangled states~(\ref{m39},\ref{m41}) that one may associate a specific CFT to some of such states.

\section{Projections on Holographic Subspaces}
The Hilbert space we obtained is still much too large to allow for a holographic interpretation. At each genus, a one-point
function of an arbitrary 
 CFT with continuous spectrum and obeying the bound~(\ref{m55}) gives rise to a vector $\mathfrak{V}$ 
in the Hilbert space defined in eq.~(\ref{m41}). So, at each genus $g$, a vector $v\in V_{\lambda,\lambda'}$ 
appears with multiplicity
 given by the dimension of the Hilbert space $H^g_{\lambda,\lambda'}$.  The sum over genera gives infinite dimensionality
 for each vector in each of the irreducible representations $V_{\lambda,\lambda'}$. So, even if on a given surface $\S_g$ 
 one could project the Hilbert space $H^g_{\lambda,\lambda'}$ to the subspace spanned by VEVs 
 $\langle O_{\lambda,\lambda'} \rangle$ belonging to a single CFT, the sum over genera would create an 
 infinite multiplicity of states. 
 
 One possibility to select a space where each $v\in H_{\lambda,\lambda'}\equiv \sum_g H^g_{\lambda,\lambda'}$ 
 appears with finite multiplicity is to consider topological
 transitions in pure gravity. They can be computed in the functional integral approach to quantum gravity.
 
 Since the gauge group is a product we can compute transition amplitudes by computing the functional integral first 
 separately for each $\slr$ gauge group and then multiplying together the results.\footnote{As in the rest of this section
 we are being a bit cavalier here, since we ignore possible boundary terms that mix the two $\slr$.}
 Because the spaces we are considering have universal asymptotic $AdS_3$ regions with conformal boundaries, we 
 must first disentangle those regions, which are black hole exteriors, from the interior regions. On the surface $\S$ such
 disentangling induces the decomposition shown in figure~(\ref{collar}). The decomposition gives a collar plus a region 
 with a fixed holonomy $2\cosh 2\pi \lambda =\Tr P \exp \int_{\partial\S} A $ on the boundary. To find a topological transition we must
 compute a functional integral on a 3D space $M$ with initial data given by a fixed flat $\slr$ connection on a surface 
 $\S^I_g$ of genus $g$ and final data given by another fixed $\slr$ flat connection on a surface $\S^F_{g'}$ of genus $g'$. 
 
 Can a transition occur between surfaces of different genus? The simplest possibility is that the transition is mediated by
 a complex saddle point, i.e. that there exists a classical solution of the CS equations of motion for the complexified 
 gauge group $\slc$, {\em which belongs to a steepest descent path in the same homology class as the path 
 $A\in \slr$}~\cite{w-ana}. 
 
\begin{figure}[h]
\begin{center}
\epsfig{file=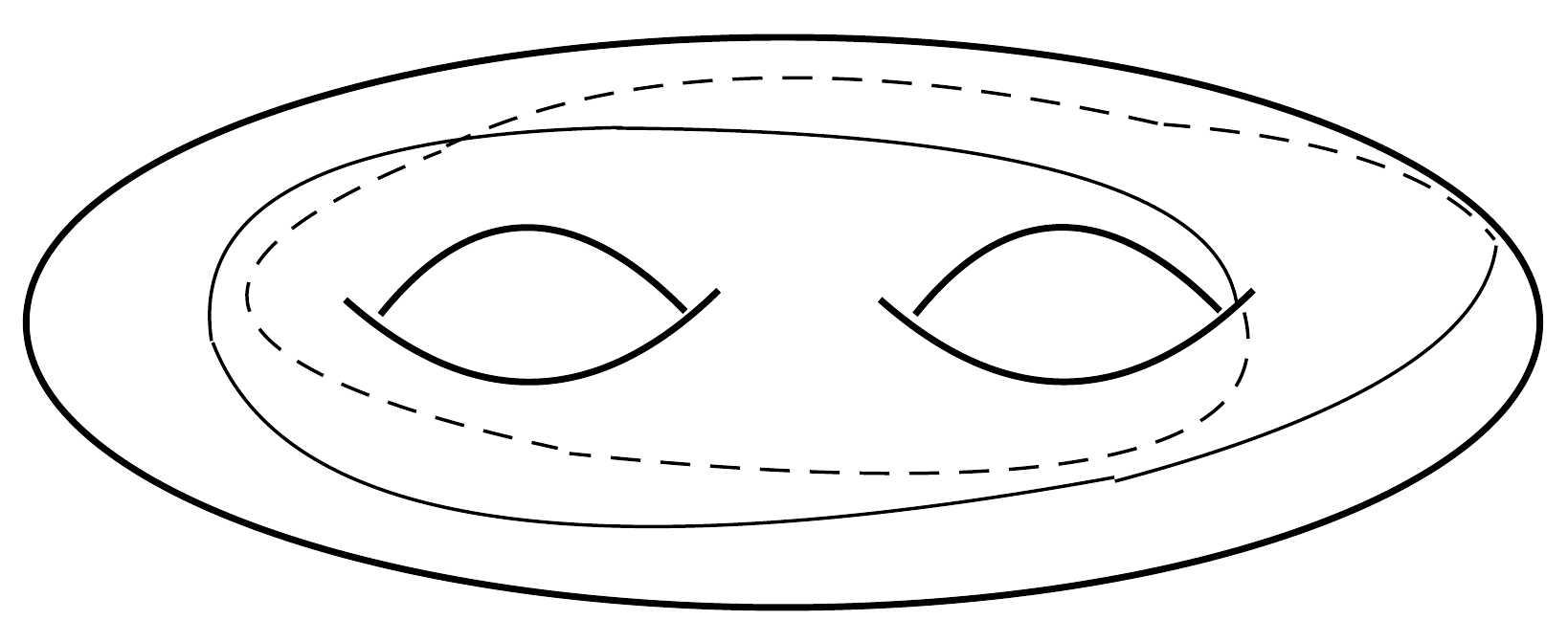, height=1.5in, width=3.5in}
\end{center}
\caption{A homotopy cycle, non-contractible in the 3D bulk, that partitions a double torus $g=2$ into two one-hole tori.}
\label{transition}
\end{figure}

 A classical solution of CS equations is a  flat $\slc$ connection in $M$. The boundary of $M$ is $\partial M=\S^I_g \cup
 \S^F_{g'}$. The boundaries of the two surfaces, which represent only the interior of a black hole, 
 are glued together to form a closed surface; at their common boundary 
 $\partial \S^I_g = \partial \S^F_{g'}$ the holonomy of the gauge field is fixed. Since the classical connection is flat in $M$,
 the holonomy is fixed also inside $M$. We arrive thus at a topological obstruction for the existence of a classical solution
 on $M$: it must be possible to decompose $\partial M$ in two surfaces $\S^I_g,\S^F_{g'}$ along a nontrivial 
 homotopy cycle in $\partial M$ that is not contractible in $M$. An example of such a homotopy cycle 
 for $g=g'=1$ is given in figure~(\ref{transition}). 

 The topological obstruction for AdS gravity is weaker than for gravity at zero cosmological constant. In the latter case one is 
 interested in transitions between closed surfaces $\S^I$, $\S^F$~\cite{w89}. The functional integral that defines the topological 
 transition is performed
 on $ISO(2,1)$ gauge fields defined on a 3-manifold $M$ such that $\partial M=\Sigma^I\cup \Sigma^F$. Functional integration over
 the dreibein $e^a$ makes the $SO(2,1)$ gauge fields flat {\em off shell}~\cite{w89} on $M$. The initial state in the transition is defined 
 by assigning $SO(2,1)$ holonomies around all non-contractible homotopy cycles  of $\S^I$; likewise, the final state is defined by
 assigning $SO(2,1)$ holonomies around all non-contractible homotopy cycles of $\S^F$. Since the $SO(2,1)$ gauge field is flat 
 off shell and the $SO(2,1)$ flat connection must have maximum Euler class to represent a nonsingular geometry~\cite{w}, several 
 topological selection rules follow. In particular, when a homotopy cycle on $\S^I$ can be deformed in $M$ to a homotopy cycle
 on $\S^F$, then the $SO(2,1)$ holonomy must be the same on $\S^F$ as on $\S^I$~\cite{w89}. 
 Moreover, no homotopy cycle can be
 contractible in $M$~\cite{ah}, since all $SO(2,1)$ holonomies must be hyperbolic. 
 
 In $\slr\times\slr$ Chern-Simons  theory one gives instead ``Verlinde" boundary conditions, i.e. one fixes the $\slr$ 
 gauge field components 
 $e^+_z,e^+_{\bar{z}},\omega_z$ on $\S^I_g$. On $\S^F_{g'}$ one likewise fixes the components 
 $e^-_z,e^-_{\bar{z}},\omega_{\bar{z}}$. These are also known in CS literature as ``Nahm pole'' boundary conditions~\cite{gawi}; 
 since they specify only half of the gauge field components on $\partial M$, they are insufficient to specify completely
 the holonomies of the gauge fields on either $\S^I_g$ or $\S^F_{g'}$. 
 
Formally the computation thus starts from the functional integral
\bea
K[B^I|B^F]&=&\int_R [dA] \exp iI_{CS}, \qquad R=\{ A\in \slc | A \in \slr\}  , \nonumber \\
A^V|_{\S^I_g} &=& B^I, \qquad A^V|_{\S^F_{g'}}=B^F .
\eea{m57}
The notation $A^V$ in this equation is a shorthand to signify that only half of the $\slr$ gauge connection components 
are specified by the boundary conditions. 
Here we must point out  a well-known but  important fact; namely, the role of reality conditions on the gauge field. To be explicit, write
the gauge field in an $SU(1,1)$ basis, as before our eq.~(\ref{constr}) in the introduction. The gauge field components then
 obey the reality conditions $\overline{e^+}=e^-$, $\overline{\omega}=\omega$. So it seems that the whole gauge field 
 $A|_{\partial M}$ is fixed by the boundary conditions, not just half of its components; moreover, 
 the gauge connection is not flat for 
 generic values of $B^I,B^F$. This means simply that a flat connection, that is a solution to the classical field equations, 
 generically exists only for connections taking value in the complexified algebra $\slc$ {\em even on the boundary $\partial M$}. A flat
 $SU(1,1)$ connection exists only for initial and final data that can be cast in the form~(\ref{m17},\ref{m7}). For generic values of 
 $B^I,B^F$, instead, 
 the functional $K[B^I|B^F]$ obeys the Gauss law constraints ${\cal G}^\pm K=0$, ${\cal G}^0 K=0$ as functional
 equations in which, as in subsection 1.3, $k e^-_{\bar{z}} =-i {\delta /  \delta e^+_z}$ and so on. 
 
In a semiclassical computation of the functional integral, 
the contour of integration is then deformed to an equivalent steepest descent contour, which is in general a union of 
Lefshetz thimbles~\cite{w-ana}, i.e. steepest descent contours starting from stationary points 
$\delta I/\delta A=0$, $A\in \slc$. 

The thimbles are determined by a Morse function~\cite{w-ana} 
\beq
M=iI_{CS}-i\bar{I}_{CS} +i\int_{\partial M} \Tr A \wedge * A.
\eeq{morse}
This definition differs from the Morse function in~\cite{w-ana} by a boundary term, 
which does not contribute to the variational equations that determine the thimbles. The variation 
$\delta M= i \int \Tr \delta A \wedge F -i \int \Tr \delta \bar{A} \wedge \bar{F}$ defines also the downward flow 
\beq
{d A \over dt} = i * \bar{F}.
\eeq{m58}
For a fixed point of $I_{CS}$ to contribute to the functional integral there must exist a downward flow starting from 
$A\in \slr$ and ending on the fixed point~\cite{w-ana}. 

At the point $A\in \slr$ one has obviously $M=0$ so the downward flow leads to a point with $M<0$. When one writes a 
connection in $\slc$ as $A=\omega +ie$ the Morse function $M$ becomes the action for Euclidean $\Lambda <0$
gravity modulo possible boundary terms. When computed on a solution of the classical equations of motion for the complexified gauge field $A=\omega + ie \in\slc$, $M$ reduces to
\beq
M= -{8\over 3} \int_M \Tr e\wedge e\wedge e .
\eeq{m59}
Our definition~(\ref{morse}) removes a  boundary term of indefinite sign arising from the on-shell value of $iI_{CS}-i\bar{I}_{CS}$,
 so one obtains that
\begin{quotation}
Whenever the stationary point of $M$ can be interpreted as a solution of pure $AdS_3$ gravity, $M<0$.
\end{quotation}
So, ``good" stationary points, that admit a Euclidean metric interpretation, 
are precisely those that are joined to the $A\in \slr$ path of integration by 
a downward flow and hence can contribute to the path integral. 

The saddle point approximation for integral~(\ref{m57}) is 
\beq
K[B^I|B^F]= \exp iI_{CS} = \exp \left(-{l\over 12\pi G}\int_M \Tr e\wedge e\wedge e +... \right) \exp (i\Re I_{CS}).
\eeq{m60}
The volume element $\int_M \Tr e\wedge e\wedge e $ and the boundary terms in $\Re i I_{CS}$, denoted by $...$ 
in~(\ref{m60}) has been shown to be proportional to the Liouville action $S_{ZTL}$ given by~(\ref{m24}) in~\cite{k}, 
in the case of closed Riemann surfaces that are Schottky doubles of open surfaces. To
establish proportionality of the actions in our case, we would need a careful treatment of boundary terms and volume 
divergences. This has been done in~\cite{k} for the spaces studied there. Here we will simply assume that the results 
of~\cite{k} hold also for our class of Riemann surfaces, so that $2\Re iI_{CS}= cS_{ZTL}$, $c=3l/2G$. The action 
$S_{ZTL}$ requires a choice of of homology basis (though it is independent of it). One of the homology cycles partitions
$\partial M$ into $\S^I_g$ and $\S^F_{g'}$; on that cycle the holonomy is fixed to be
$\Tr P \exp \int_{\partial\S^I_{g}} A = 2\cosh 2\pi \lambda $. 

$AdS_3$ gravity is given by the $\slr \times \slr$ CS theory, so we need to repeat the derivation leading to 
eq.~(\ref{m60}) for the other $\slr$ factor. The leading semiclassical action one can get is of course obtained by 
choosing for each $\slr$ factor the classical solution with the largest value of $\Re iI$. One can choose the classical
saddle point actions in the two $\slr$ such that the actions on the first factor $I^1_{CS}$ and the action on the
other $I^2_{CS}$ obey $I^1_{CS}=-\bar{I}^2_{CS}$. So, in the semiclassical approximation, the functional integral 
of quantum gravity reduces  $\exp (iI -i\bar{I})=\exp cS_{ZTL}$. 

The exponential of the Liouville action evaluated on-shell is 
itself the semiclassical approximation to the partition function of the Liouville CFT $Z_{CFT}(\partial M)$. On the surface 
$\partial M=\S^I_g\cup \S^F_{g'}$  $Z_{CFT}$ factorizes because of general formulas of CFTs as 
\beq
Z_{CFT}(\S^I_g\cup \S^F_{g'})\sim \int d\mu \langle V^\mu \rangle_{\S^I_g}\langle V^\mu \rangle_{\S^F_{g'}}.
\eeq{m61}
where in general the integral $d\mu$ is over all primaries and descendants of the CFT.
In our case, since one of the holonomies is fixed to $\mu=\lambda$, the partition function $Z(\S^I_g\cup \S^F_{g'})$ 
should instead factorize into the product 
\beq
Z_{CFT}=C(g,\lambda)C(g',\lambda)\langle V^\lambda \rangle_{\S^I_g}\langle V^\lambda \rangle_{\S^F_{g'}},
\eeq{m62}
where $V^\lambda$ is a primary vertex with conformal weight $\Delta(\lambda)$ and $C(g,\lambda)$ is a constant 
that could be determined by a more accurate computation. 
This derivation is of course at best heuristic but we shall adopt it and defer a better derivation of $K[B^I|B^F]$ to future work. 

Equation~(\ref{m62}) has one intriguing property: 
it projects the Hilbert space $\mathfrak{H}$ in~(\ref{m56a}) precisely onto the Hilbert space of Liouville CFT.
It also selects only a specific linear combination of VEVs on surfaces of different genus. 
So a given Virasoro$\times$Virasoro representation
multiplies only one vector, instead of (at least) one for each genus. We thus 
conjecture that the projected Hilbert space $\mathfrak{L}$ is spanned by vectors
\beq
\mathfrak{x}=\int_{\Delta(\lambda)
\geq (c-1)/24}d\lambda  V_{\lambda,\lambda}\times \sum_{g=1}^\infty C(g,\lambda)\langle V^\lambda\rangle_g .
\eeq{m63}
Comparing with eq.~(\ref{m56a}) we notice that the infinite multiplicity of Virasoro representations has been replaced
by a single vector. The vacuum Virasoro representation has also disappeared. 

Evidently, the derivation leading to eq.~(\ref{m63}) ought to be made more precise. Among other shortcuts, we paid little attention to
the precise relation between the kernel $K[B^I | B^F]$ and  the equivalent operator defined ``downstairs'' on the space 
$\mathfrak{H}$ given by eq.~(\ref{m56a}). Specifically, in going from eq.~(\ref{m60}) to eq.~(\ref{m61}), 
we skipped several important points,
such as whether the VEV is given by eq.~(\ref{m39}) or eq.~(\ref{m44}), or whether one needs to multiply $K[B^I | B^F]$ by the
prefactor $H$ in~(\ref{m29}). We hope to come back to these points in a future work, finding for the time being 
some comfort in 
realizing that eq.~(\ref{m62}) does define an interesting and sensible projection. One subtlety that does not show up in AdS
gravity is the infrared divergence~\cite{w89} associated with zero modes of the dreibein. In $ISO(2,1)$ Chern-Simons gravity, 
infrared divergences may ultimately suppress topology-changing amplitudes~\cite{caco}. In AdS gravity, at least for the dominant 
saddle point --where $M$ is a handlebody and there is a unique flat gauge field that obeys the boundary conditions on $\S^I_g\cup
\S^F_{g'}$-- there are no zero modes that may originate an infrared divergence.

\section{Scattered Last Remarks}
Both the large Hilbert space~(\ref{m56a}) and the small, projected one~(\ref{m63}) are problematic for a fully consistent 
theory of quantum gravity. 

The large Hilbert space cannot be obtained in any single CFT; in fact it is the target space for a large 
class of them. It is nevertheless too small to contain in any precise sense CFTs that are usually associated with 
holographic duals of quantum gravity. By this we mean the following. A VEV of a vertex in a CFT defines a vector in 
$\mathfrak{H}$ (actually a Virasoro$\times$Virasoro irrep) only if the vertex belongs to a CFT with continuous spectrum and no 
$\slc$-invariant vacuum. The range of conformal weights for the vertex is also that associated to normalizable
{\em states} in the CFT, so it excludes conformal weights associated to non-normalizable operators. 
On the other hand, if a CFT were to reproduce basic properties of quantum gravity, such as the Bekenstein-Hawking 
formula for black holes entropy~\cite{strom}, it must have discrete spectrum --lest the entropy of any finite-energy state is
 infinite-- and an $\slc$ 
invariant vacuum --otherwise $c_{eff}<c$ in Cardy's~\cite{cardy} formula~\cite{car} $S=2\pi \sqrt{c_{eff}\Delta/6} + 
2\pi\sqrt{c_{eff}\bar{\Delta}/6}$. 

So we could say the Hilbert space~(\ref{m56a}) is at the same time too large and too small to allow for
a quantum theory that provides a microscopic origin of black hole entropy. 

The projected Hilbert space $\mathfrak{L}$ at
least solves the first problem. In it, indeed, each irrep of the Virasoro$\times$Virasoro algebra is multiplied by a unique linear 
combination of VEVs, on surfaces of different genera, of the (unique) operator of weights $(\lambda,\lambda)$ in 
Liouville conformal field theory. 

Still, since in Liouville CFT the lowest  dimension of a state is $\Delta=(c-1)/24$ and $c_{eff}=c-24\Delta$, the
theory has the same effective central charge as a single noncompact free scalar: $c_{eff}=1$ (after all, in many respects 
Liouville theory is a free scalar in heavy disguise). Moreover, the problem remains that the spectrum of states in 
$\mathfrak{L}$ is continuous, so properly speaking all black holes have infinite entropy. Finally, all Liouville primaries
have $\Delta=\bar{\Delta}$ so they correspond only to non-rotating BTZ black holes. 

The latter property may hint at the presence  of some extra heavy states besides black holes in a 
UV complete theory of pure gravity. 
Specifically, long strings with tension tuned to be equal to a two-form charge\footnote{A gauge invariant two-form carries 
no degrees of freedom in 3D.} can escape to the boundary of $AdS_3$ and so have continuous spectrum. Such additional
states  were proposed in~\cite{m-w} to account for the lack of holomorphic factorization and other problems of the torus 
partition function of Euclidean $\Lambda<0$ pure gravity in 3D. In~\cite{kp1,kp2} the classical dynamics of long string 
was studied beyond the probe approximation. It was shown that in the limit that their tension is Planckian, $T\sim G^{-2}$, 
their collapse generate only heavy states
with any mass above the ``Seiberg bound'' $M=(c-1)/12l=1/8G-1/12l$. So, it is at least possible to modify pure gravity 
without introducing any sub-Planckian degree of freedom, in such a  manner as to produce all and only the states found in 
Liouville theory.  Of course, the existence of heavy strings changes pure gravity at Planckian energies. It also 
introduces a process whereby the
cosmological constant changes by the Brown-Teitelboim mechanism~\cite{bt,kp1,kp2}. 

On the other hand, the fact that $c_{eff}<c$ may not be a liability after all for a theory of pure gravity, but rather an unavoidable
 feature. It was indeed argued in~\cite{gkrs}, that theories where black holes are not dynamical, as it is the case in pure gravity 
 with or without heavy strings, $c_{eff}$ must be strictly less than $c$. Explicit calculations, in a linear dilaton background and in
 strings on $AdS_3$ at $l_S>l$, support this conjecture~\cite{gkrs}. The proposal of~\cite{gkrs} fits beautifully with
 our findings: when $c=c_{eff}$ gravity is in a phase where the high-energy spectrum is dominated by black holes, that 
 must perforce be dynamical states. When $c_{eff}<c$, other states, long critical strings perhaps, 
 dominate the high-energy spectrum, because black holes are not dynamical objects. Pure gravity, with $c_{eff}=1$ 
 is in this interpretation just an extreme case of the latter possibility.
 
 There is one case where, perhaps, canonical quantization could be compatible with dynamical black holes. This is 
 when $c<1$. Of course this is also when we cannot compare our results with semiclassical expectations, because 
 the curvature of $AdS_3$ is Planckian ($l<2G/3$). It is also the case where different formulas for the measure of 
 integration on \tch\ space, which coincide at large $c$ but differ by subdominant terms in the $1/c$ expansion, produce 
 vastly different results. For lack of a better reason, we will more or less arbitrarily keep definition~(\ref{m54}). This
 definition gives bound~(\ref{m55}) $\Delta\geq (c-1)/24$. For $c<1$, the $\slc$-invariant vacuum is then normalizable, 
 so the first obstacle to a conventional CFT interpretation of the Hilbert space is absent.

 The second obstacle is that
 the states flowing inside conformal blocks belong to the continuous spectrum of $L_0,\bar{L}_0$. This property 
 seems unavoidable, because the space $[T(\S)\times T'(\S)] / M$ is noncompact. This seems also in blatant contradiction
 with the fact that if we were able to project the Hilbert space $\mathfrak{H}$ on a single CFT with $c<1$, such theory would
 necessarily belong to the minimal {\em unitary} series~\cite{fqs}, $c=1-{6\over (m+1)m}$. Minimal CFTs contain a finite
 number of Virasoro primaries so their spectrum is discrete. 
 
 A possible way out of this contradiction is found in another property of minimal CFTs. 
 In~\cite{bant} it was proven that on a torus all VEVs of 
 primary operators in a $c<1$ minimal model are invariant under a {\em normal, finite-index} subgroup $\cal N$ of the 
 modular group. So, in view of factorization~(\ref{m56}), 
 at least for $\S=$ a one-holed torus, one can restrict the range 
 of integration in e.g.~(\ref{m46}) from $[T(\S)\times T'(\S)] / M$ to
 \beq
  {{T(\S)\over {\cal N}} \times {T'(\S) \over {\cal N}} \over M},
 \eeq{m64}
 Since $\cal N$ is normal and finite-order, $T(\S)/{\cal N}$ is compact; its volume is a finite, integer multiple of the volume of
 moduli space. Normalizability of the wave function now {\em implies} that the spectrum of conformal weights in the 
 conformal blocks is discrete, as it should. Factoring out $\cal N$ can be seen as resulting from imposing a new, nonclassical gauge 
 symmetry, invisible in the large-$c$ limit. This construction and interpretation is reminiscent of ref.~\cite{ising}, which also
 argued for new gauge symmetries when studying the gravity dual of the Ising model. In~\cite{ising} the agreement 
 between bulk and boundary computations worked best for the $c=1/2$ case. Here we do not see a difference between
 $c=1/2$ and other minimal models.... but of course our conjecture is at best only a guess. 
 To make it work, one would need a symmetry
 equivalent to $\cal N$  also for higher-genus surfaces --i.e. one wold need a normal, finite order subgroup of the 
 mapping class group of $\S_g$ for any $g$. We do not know if such symmetry exists. 
 
 Even though canonical quantization of $c<1$ pure gravity seems consistent with the holographic $AdS_3/CFT_2$ duality, 
 it applies, as we said, to a regime 
 far outside the semiclassical one. It would be interesting to extend our study to higher Chern-Simons theories and 
 higher-spin $AdS_3$ theories. An equivalent of the Brown-Henneaux result exists in such cases~\cite{CFPT}; therefore, 
  it may be possible
 to attain a controllable semiclassical regime in some of those theories. One concrete possibility is offered by  
 theories dual to $W_N$ minimal models in the scaling limit studied extensively by Gaberdiel and Gopakumar~\cite{gg}.
 \subsection*{Acknowledgements}
 We would like to thank L. Alvarez-Gaum\'e, R. Benedetti, K. Krasnov, J. Maldacena, A. Maloney, R. Mazzeo, P. Menotti, J. Teschner
  and E. Witten for useful 
 discussions and suggestions accumulated over the course of the several years that this paper took to materialize. 
 Thanks to them the wall of our
 ignorance now shows some cracks. This paper started to be written in the present form while one of us (M.P.) was visiting 
 the lnstitute for Advanced Study, Princeton, that we thank for hospitality.  Research supported in part by
 NSF grant PHY-1316452. 
\setcounter{section}{0}
\renewcommand{\thesection}{\Alph{section}}
\section{Volume of $T_{(1,1)}$ Moduli Space}
\setcounter{equation}{0}
\renewcommand{\theequation}{A.\arabic{equation}} 
To check that we found a correct symplectic form in section 3, let us compute the volume of the {\em moduli space} 
of a one-hole torus.

For this purpose it is more convenient to introduce new coordinates following~\cite{wol83}.
\beq
a = \frac{x}{yz}, \quad b = \frac{y}{zx}, \quad c = \frac{z}{xy}.
\eeq{abc_coord}
In these coordinates the \emph{rough fundamental domain} of the moduli space is the region enclosed by three curves
\beq
a= \frac{1}{2}, \quad b = \frac{1}{2}, \quad \mathcal{C} = \{ \left(a,b \right) \vert a + b + 2 \sinh\left(\lam / 2 \right) a b - \frac{1}{2} =0 \}.
\eeq{region2}
A rough fundamental domain for the moduli space ${\cal M}(\S)$ is a region of \tch\ space $T(\S)$ defined as follows: 
${\cal M}(\S)$  is obtained by quotienting $T(\S)$ by the mapping class group of the Riemann surface $\S$. It can be 
identified with the fundamental domain in $T(\S)$ under the action of the mapping class group. A rough fundamental 
domain is a region of $T(\S)$   that contains at most a finite number of
images under the mapping class group of any point in the fundamental domain.  In our case, a rough fundamental 
domain can be obtained from the 
Teichm\"uller space by the action of $\Gamma(2)$, the level-2 principal congruence subgroup of the modular group 
${M} \approx SL(2, \mathbb{Z})$. $\Gamma(2)$ is an index-$6$ subgroup of the mapping class group~\cite{keen, wol83}.
By expressing the symplectic form in terms of $a$ and $b$, the volume is given by the integral expression
\bea
V &=& \int_{\td{\mathcal{F}}} { \frac{da \wedge db}{ab\left( 1 - a - b \right)} } \nonumber \\
&=& \int^{1/2}_0 db{ \int^{1/2}_{\mathcal{C}} { da \frac{1}{ab\left(1-a-b\right)} } } \nonumber \\
&=& -4 \int^1_0 { \frac{dy}{1-y^2} \log y }  + 2 \int^1_0 { \frac{dy}{1-y^2} \log \left[ \cosh^2 \left( \lam / 2 \right) -  y^2 \sinh^2 \left(\lam/2\right) \right] } .
\eea{vol}
Despite their appearance, both integrals in the last line converge because the numerator in both expressions
vanishes fast enough at $y=1$ to make the integrand regular at $y=1$. 
In particular,  the second integral equals $\lam^2 /2 = L^2 / 8$, where $L$ is the length of the geodesic boundary.
After accounting for the index of the subgroup, the volume of the moduli space of a one-hole torus  is
\beq
V = \frac{\p^2}{6} + \frac{L^2}{24} ,
\eeq{vol2}
which agrees with the result given in \cite{mirz}.
Notice that the volume of the moduli space of a torus with a hole is the sum of the volume of the moduli space of a 
torus with a puncture, see \cite{wol83}, and a polynomial in the length of the boundary.\footnote{A generalization to a 
genus $g$ Riemann surface with $n$ geodesic boundaries is given in \cite{mirz}.}
This result is rooted in the fact that the symplectic form of a Riemann surface with a boundary is cohomologous to the 
sum of the symplectic form of once punctured Riemann surface plus the first Chern class of the tangent bundle at the 
puncture~\cite{mirz}. We gave a detailed discussion on this point and its physical implications in section 5.

\section{From Fricke-Klein to Fock coordinates}
\setcounter{equation}{0}
\renewcommand{\theequation}{B.\arabic{equation}}
In the study of quantum 
Teichm\"uller theory, Fock coordinates are useful~\cite{fock} because they define canonical variables.
So it is worth to present the explicit transformation between the Fricke-Klein coordinates and the Fock coordinates in the
case of a one-boundary torus.

Once we are given  a Riemann surface, Fock coordinates can be obtained by drawing a fat graph of it.
For the one-boundary torus the fat graph has three edges and we assign a real number to each edge, which becomes a 
component of the Fock coordinate system.
By calling these coordinates  $\mathfrak{a}$, $\mathfrak{b}$ and $\mathfrak{c}$, subject to the constraint 
$\mathfrak{a}+ \mathfrak{b} + \mathfrak{c}=\lam$ and using their relation to length operators 
[see for example~\cite{terayama}], we obtain
\beq
x=2\cosh\left(\tfrac{\lam-\mathfrak{b}}{2}\right)+e^{\mathfrak{a}} e^{-\frac{\lam-\mathfrak{b}}{2}} , \quad y=2\cosh\left(\tfrac{\lam-\mathfrak{a}}{2}\right)+e^{-\mathfrak{a}} e\mathfrak{b}^{\frac{\lam-\mathfrak{a}}{2}} .
\eeq{focktofri}
This enables us to reduce the symplectic form (\ref{symp_xyz}) to $d \mathfrak{a} \wedge d \mathfrak{b}$ 
up to a constant factor.

To find the inverse of transformation~(\ref{focktofri}), i.e. a transformation from Fricke-Klein coordinates to Fock 
coordinates, it is convenient to define $\mathfrak{A} = e^{\mathfrak{a} /2}$, $\mathfrak{B} = e^{\mathfrak{b} /2}$ and 
$\Lambda = e^{\lam /2}$ and express $x$ and $y$ in terms of 
a quadratic equations in $\mathfrak{A}$, $\mathfrak{B}$ and $\Lambda$.
Then $\mathfrak{A}$ can be expressed in terms of $x$ and $y$ by eliminating $\mathfrak{B}$, and $\mathfrak{B}$ can be obtained using the symmetry of the coordinate transformation (\ref{focktofri}); $x \leftrightarrow y$ under $\mathfrak{a} \leftrightarrow -\mathfrak{b}$ and $\lam \leftrightarrow -\lam$ :
\bea
\mathfrak{a}&=&2 \log \left[ y \{ \left( x^2 -4 \right) + 4 e^{\lam/2} \cosh\left(\lam/2\right) \} \pm x \sqrt{\left( x^2 -4\right)\left(y^2 -4\right)-16\cosh^2\left(\lam/2\right)}\right] \nonumber \\
&\qquad& - 2\log\left[2\left(x^2 e^{-\lam/2} + y^2 e^{\lam/2}\right)\right] , \nonumber \\
\mathfrak{b}&=&-2\log\left[x\{\left(y^2 - 4 \right) + 4 e^{-\lam/2} \cosh\left(\lam/2\right) \} \pm y\sqrt{\left(x^2 -4\right)\left(y^2 -4\right)-16 \cosh\left(\lam/2\right)}\right] \nonumber \\ 
&\qquad& + 2\log\left[2\left(x^2 e^{-\lam/2} + y^2 e^{\lam/2}\right)\right] .
\eea{fritofock}
By simplifying these coordinates transformations by setting $\lam=0$, which corresponds to the case of a torus with a puncture, one can explicitly check that the symplectic form $d \mathfrak{a} \wedge d \mathfrak{b}$ reproduces the symplectic form (\ref{symp_xyz}) in Fricke-Klein coordinates.
The transformation mapping our new coordinates to the Fock coordinates is obtained by using eqs.~(\ref{fricke}).


\begin{thebibliography}{999}

\bibitem{wdw} 
B.S. DeWitt, ``Quantum Theory of Gravity. I. The Canonical Theory," Phys.\ Rev. {\bf 160}, 1113 (1967).

\bibitem{adm}
F.~A.~E.~Pirani and A.~Schild, ``On the Quantization of Einstein's Gravitational Field Equations,'' Phys.\ Rev.\ {\bf 79}, 986 (1950);
P.~A.~M.~Dirac, ``Fixation of coordinates in the Hamiltonian theory of gravitation,'' Phys.\ Rev.\ {\bf 114}, 924 (1959); 
R.~L.~Arnowitt, S.~Deser and C.~W.~Misner, ``Canonical variables for general relativity,'' 
Phys.\ Rev.\ {\bf 117}, 1595 (1960).

\bibitem{books} 
  T.~Thiemann,
  {\em Modern canonical quantum general relativity},
  Cambridge, UK: Cambridge Univ. Pr. (2007) 819 p
  [gr-qc/0110034]; C.~Rovelli,
  {\em Quantum gravity},
  Cambridge, UK: Univ. Pr. (2004) 455 p  
  
  \bibitem{rov}
  C.~Rovelli, ``Notes for a brief history of quantum gravity,''
  gr-qc/0006061;
  B.~DeWitt,
  ``Quantum Gravity Yesterday and Today,''
  Gen.\ Rel.\ Grav.\  {\bf 41}, 413 (2009)
  [Gen.\ Rel.\ Grav.\  {\bf 41}, 671 (2009)]
  [arXiv:0805.2935 [physics.hist-ph]].  
  
  \bibitem{in-out}
  H.~Nicolai, K.~Peeters and M.~Zamaklar,
  ``Loop quantum gravity: An Outside view,''
  Class.\ Quant.\ Grav.\  {\bf 22}, R193 (2005)
  [hep-th/0501114];  
  T.~Thiemann,
  ``Loop Quantum Gravity: An Inside View,''
  Lect.\ Notes Phys.\  {\bf 721}, 185 (2007)
  [hep-th/0608210].  
  
  
  \bibitem{at} A.~Achucarro and P.~K.~Townsend,
  ``A Chern-Simons Action for Three-Dimensional anti-De Sitter Supergravity Theories,''
  Phys.\ Lett.\ B {\bf 180}, 89 (1986).
  
  \bibitem{w} E.~Witten,
  ``(2+1)-Dimensional Gravity as an Exactly Soluble System,''
  Nucl.\ Phys.\ B {\bf 311}, 46 (1988).
  
 \bibitem{BTZ}   
M. Ba\~nados, C. Teitelboim and J. Zanelli, ``Black Hole in Three-Dimensional 
Spacetime,'' Phys. Rev. Lett. \textbf{69}, 1849 (1992);
M. Ba\~nados, M. Henneaux, C. Teitelboim and J. Zanelli, ``Geometry of the (2+1) black hole,''
  Phys.\ Rev.\ D {\bf 48}, 1506 (1993)
  [Phys.\ Rev.\ D {\bf 88}, no. 6, 069902 (2013)]
  [gr-qc/9302012].
  
  \bibitem{V}
H. Verlinde, ``Conformal Field Theory, Two-dimensional Quantum Gravity and Quantization of 
Teichmuller Space," Nucl. Phys. B337(1990), 632-680.

 \bibitem{VV}
H. Verlinde, E. Verlinde, ``Conformal Field Theory and Geometric Quantization,'' PUPT-89/1149.  

\bibitem{deser}
S.~Deser, R.~Jackiw and S.~Templeton,
  ``Topologically Massive Gauge Theories,''
  Annals Phys.\  {\bf 140}, 372 (1982)
  [Annals Phys.\  {\bf 185}, 406 (1988)]
  [Annals Phys.\  {\bf 281}, 409 (2000)];
  ``Three-Dimensional Massive Gauge Theories,''
  Phys.\ Rev.\ Lett.\  {\bf 48}, 975 (1982).  
  
  \bibitem{Wit}
E. Witten, ``Quantum Field Theory and the Jones Polynomial," Commun. Math. Phys. 121(1989), 351-399

 \bibitem{MS}
S. Elitzur, G. Moore, A. Schwimmer, N. Seiberg, ``Remarks on the 
Canonical Quantization of the Chern-Simons-Witten Theory,'' Nucl. Phys. B326(1989) 108-134.

\bibitem{far-kra}
H.~M. Farkas and I. Kra, {\em Riemann surfaces,} Springer New York, 1992.

\bibitem{cb}
A.~A.~Belavin, A.~M.~Polyakov and A.~B.~Zamolodchikov,
  ``Infinite Conformal Symmetry in Two-Dimensional Quantum Field Theory,''
  Nucl.\ Phys.\ B {\bf 241}, 333 (1984).
  
\bibitem{tesch}
J.~Teschner,  ``From Liouville theory to the quantum geometry of Riemann surfaces,''
  hep-th/0308031.
  
  \bibitem{tesch2}
  J.~Teschner and G.~S.~Vartanov, ``Supersymmetric gauge theories, quantization of $\mathcal{M}_{\mathrm{flat}}$, 
  and conformal field theory,'' Adv.\ Theor.\ Math.\ Phys.\ {\bf 19}, 1 (2015) [arXiv:1302.3778 [hep-th]]. 
  
  \bibitem{mod}
  J.~Teschner,
  ``Quantum Liouville theory versus quantized Teichmuller spaces,''
  Fortsch.\ Phys.\  {\bf 51}, 865 (2003)
  [hep-th/0212243];
  ``On the relation between quantum Liouville theory and the quantized Teichmuller spaces,''
  Int.\ J.\ Mod.\ Phys.\ A {\bf 19S2}, 459 (2004)
  [hep-th/0303149];  J.~Teschner,
  ``An Analog of a modular functor from quantized teichmuller theory,''
  math/0510174 [math-qa].  
  
  \bibitem{flat}
  J.~Teschner,
  ``Quantization of moduli spaces of flat connections and Liouville theory,''
  arXiv:1405.0359 [math-ph].  
  
  \bibitem{agt}
  L.~F.~Alday, D.~Gaiotto and Y.~Tachikawa,
  ``Liouville Correlation Functions from Four-dimensional Gauge Theories,''
  Lett.\ Math.\ Phys.\  {\bf 91}, 167 (2010)
  [arXiv:0906.3219 [hep-th]].  
  
  \bibitem{agt1}
  N.~Drukker, J.~Gomis, T.~Okuda and J.~Teschner,
  ``Gauge Theory Loop Operators and Liouville Theory,''
  JHEP {\bf 1002}, 057 (2010)
  [arXiv:0909.1105 [hep-th]];
  I.~Coman, M.~Gabella and J.~Teschner,
  ``Line operators in theories of class $\mathcal{S}$, quantized moduli space of flat connections, and Toda field theory,''
  arXiv:1505.05898 [hep-th].  
  
  \bibitem{agt2}
  L.~F.~Alday, D.~Gaiotto, S.~Gukov, Y.~Tachikawa and H.~Verlinde,
  ``Loop and surface operators in N=2 gauge theory and Liouville modular geometry,''
  JHEP {\bf 1001}, 113 (2010)
  [arXiv:0909.0945 [hep-th]].  
  
  \bibitem{hitch-q}
  J.~Teschner,
  ``Quantization of the Hitchin moduli spaces, Liouville theory, and the geometric Langlands correspondence I,''
  Adv.\ Theor.\ Math.\ Phys.\  {\bf 15}, 471 (2011)
  [arXiv:1005.2846 [hep-th]];
  D.~Gaiotto, G.~W.~Moore and A.~Neitzke,
  ``Spectral networks,''
  Annales Henri Poincare {\bf 14}, 1643 (2013)
  [arXiv:1204.4824 [hep-th]];
    L.~Hollands and A.~Neitzke,
  ``Spectral networks and Fenchel-Nielsen coordinates,''
  arXiv:1312.2979 [math.GT];
  
  \bibitem{shat}
  N.~Nekrasov, A.~Rosly and S.~Shatashvili,
  ``Darboux coordinates, Yang-Yang functional, and gauge theory,''
  Nucl.\ Phys.\ Proc.\ Suppl.\  {\bf 216}, 69 (2011)
  [arXiv:1103.3919 [hep-th]].  
  
  \bibitem{dpw}
  S.~Axelrod, S.~Della Pietra and E.~Witten,
  ``Geometric Quantization of {Chern-Simons} Gauge Theory,''
  J.\ Diff.\ Geom.\  {\bf 33}, 787 (1991).  
  
  \bibitem{zo-tak}
  P.~G. Zograf and L.~A. Takhtadzhyan. "On uniformization of Riemann surfaces and the Weil-Petersson metric on Teichm\"uller and Schottky spaces." Mathematics of the USSR-Sbornik 60, no. 2 (1988): 297.  
  
  \bibitem{tak-teo}
  L.~A.~Takhtajan and L.~P.~Teo,
  ``Liouville action and Weil-Petersson metric on deformation spaces, global Kleinian reciprocity and holography,''
  Commun.\ Math.\ Phys.\  {\bf 239}, 183 (2003)
  [math/0204318 [math-cv]].  
  
  \bibitem{tak-lap}
  A.~McIntyre and L.~A.~Takhtajan,
  ``Holomorphic factorization of determinants of laplacians on Riemann surfaces and a higher genus generalization of kronecker's first limit formula,''
  Analysis {\bf 16}, 1291 (2006)
  [math/0410294 [math.CV]].  
  
\bibitem{fock}
V. V. Fock, "Dual Teichm\"uller Spaces," dg-ga/9702018.
\
\bibitem{che-fo}
L.~Chekhov and V.~V.~Fock,
  ``Quantum Teichmuller space,''
  Theor.\ Math.\ Phys.\  {\bf 120}, 1245 (1999)
  [Teor.\ Mat.\ Fiz.\  {\bf 120}, 511 (1999)]
  [math/9908165 [math-qa]].  
  
  \bibitem{kash}
  R.~M.~Kashaev,
  ``Quantization of Teichmueller spaces and the quantum dilogarithm,''
  Lett.\ Math.\ Phys.\  {\bf 43}, 105 (1998).
  
  \bibitem{w8}
  E.~Witten,
  ``Three-Dimensional Gravity Revisited,''
  arXiv:0706.3359 [hep-th].    
  
  \bibitem{quill}
  D. Quillen, ``Determinants of Cauchy-Riemann operators over a Riemann surface," Functional
Analysis and Its Application {\bf 19}, 31 (1985).  

\bibitem{BH} 
J.~D. Brown and M. Henneaux, ``Central Charges in the Canonical Realization of Asymptotic Symmetries: 
An Example from Three Dimensional Gravity,'' Commun. Math. Phys. \textbf{104}, 207 (1986).

\bibitem{seg}
G.B. Segal, ``The Definition Of Conformal Field Theory,� in
U. Tillmann, ed. Topology Geometry, and Quantum Field Theory (Cambridge University Press, 2004)

\bibitem{seib}
N.~Seiberg,
  ``Notes on quantum Liouville theory and quantum gravity,''
  Prog.\ Theor.\ Phys.\ Suppl.\  {\bf 102}, 319 (1990).
  
\bibitem{tesch-liu}
J.~Teschner,
  ``Liouville theory revisited,''
  Class.\ Quant.\ Grav.\  {\bf 18}, R153 (2001)
  [hep-th/0104158].
  
\bibitem{w89}
E.~Witten,
  ``Topology Changing Amplitudes in (2+1)-Dimensional Gravity,''
  Nucl.\ Phys.\ B {\bf 323}, 113 (1989).
  
\bibitem{gkrs}
A.~Giveon, D.~Kutasov, E.~Rabinovici and A.~Sever,
  ``Phases of quantum gravity in AdS(3) and linear dilaton backgrounds,''
  Nucl.\ Phys.\ B {\bf 719}, 3 (2005)
  [hep-th/0503121].

\bibitem{bant} 
  P.~Bantay,
  ``The Kernel of the modular representation and the Galois action in RCFT,''
  Commun.\ Math.\ Phys.\  {\bf 233}, 423 (2003)
  [math/0102149].
 
 \bibitem{ising} 
  A.~Castro, M.~R.~Gaberdiel, T.~Hartman, A.~Maloney and R.~Volpato,
  ``The Gravity Dual of the Ising Model,''
  Phys.\ Rev.\ D {\bf 85}, 024032 (2012)
  [arXiv:1111.1987 [hep-th]].  
  
  \bibitem{stro1}
  W.~Li, W.~Song and A.~Strominger,
  ``Chiral Gravity in Three Dimensions,''
  JHEP {\bf 0804}, 082 (2008)
  [arXiv:0801.4566 [hep-th]].
  
  \bibitem{stro2}
  A.~Strominger,
  ``A Simple Proof of the Chiral Gravity Conjecture,''
  arXiv:0808.0506 [hep-th].  
  
  \bibitem{stro3}
  A.~Maloney, W.~Song and A.~Strominger,
  ``Chiral Gravity, Log Gravity and Extremal CFT,''
  Phys.\ Rev.\ D {\bf 81}, 064007 (2010)
  [arXiv:0903.4573 [hep-th]].  
  
  \bibitem{maloney}
   A.~Maloney, ``Geometric Microstates for the Three Dimensional Black Hole?,''
  arXiv:1508.04079 [hep-th].  
  
  \bibitem{brill}
D.~R.~Brill,
  ``Multi - black hole geometries in (2+1)-dimensional gravity,''
  Phys.\ Rev.\ D {\bf 53}, 4133 (1996)
  [gr-qc/9511022];
  S.~Aminneborg, I.~Bengtsson, D.~Brill, S.~Holst and P.~Peldan,
  ``Black holes and wormholes in (2+1)-dimensions,''
  Class.\ Quant.\ Grav.\  {\bf 15}, 627 (1998)
  [gr-qc/9707036].
  
\bibitem{barbot}
T.~Barbot,
  ``Causal properties of AdS-isometry groups. II. BTZ multi black-holes,''
  Adv.\ Theor.\ Math.\ Phys.\  {\bf 12} (2008)
  [math/0510065 [math-gt]].

\bibitem{ks}
C.~Scarinci and K.~Krasnov,
  ``The universal phase space of $AdS_3$ gravity,''
  Commun.\ Math.\ Phys.\  {\bf 322}, 167 (2013)
  [arXiv:1111.6507 [hep-th]].
  
\bibitem{monc}
V.~Moncrief,
  ``Reduction of the Einstein equations in (2+1)-dimensions to a Hamiltonian system over Teichmuller space,''
  J.\ Math.\ Phys.\  {\bf 30}, 2907 (1989).
  
  \bibitem{mess}
  G.~Mess, ``Lorentz spacetimes of constant curvature,"  Geom. \ Dedicata {]\bf 126}, 345 (2007).
  
  \bibitem{gold}
  W.~M.~Goldman, ``The symplectic nature of fundamental groups of surfaces," Adv. in Math. \ {\bf 54}, 200 (1984).
    
  \bibitem{hitch}
  N.~J. Hitchin,  ``The self-duality equations on a Riemann surface." Proc. \ London Math. \ Soc.\ {\bf 55.3}, 59 (1987).
  
  \bibitem{matsch}
  H.~J.~Matschull,  ``On the relation between (2+1) Einstein gravity and Chern-Simons theory,''
  Class.\ Quant.\ Grav.\  {\bf 16}, 2599 (1999)
  [gr-qc/9903040].  
  
  \bibitem{meus}
  C.~Meusburger,
  ``Geometrical (2+1)-gravity and the Chern-Simons formulation: Grafting, Dehn twists, Wilson loop observables and the cosmological constant,''
  Commun.\ Math.\ Phys.\  {\bf 273}, 705 (2007)
  [gr-qc/0607121]. 
  
  \bibitem{p&s}
  A. Pressley and G.~B. Segal, {\em  Loop groups},  Clarendon Press (1986). 
  
\bibitem{w_bos} 
  E.~Witten,
  ``Nonabelian Bosonization in Two-Dimensions,''
  Commun.\ Math.\ Phys.\  {\bf 92}, 455 (1984).
 
 \bibitem{fk}
 R. Fricke and F. Klein, {\em Vorlesungen uber die Theorie der Automorphen Funktionen},' Leipzig, 1926; Teubner, 1965.  
 
\bibitem{keen}
L.~Keen, 
  ``A rough fundamental domain for Teichmuller spaces,'' 
  Bull.\ Amer.\ Math. Soc.\ {\bf 83}, 1199-1226 (1977).

\bibitem{wol83}
 	S.~Wolpert, 
  ``On the K\"ahler form of the moduli space of once punctured tori,'' 
  Comment.\ Math.\ Helvetici\ {\bf 58}, 246-256 (1983). 
  
   \bibitem{chvd}
  O.~Coussaert, M.~Henneaux and P.~van Driel,
  ``The Asymptotic dynamics of three-dimensional Einstein gravity with a negative cosmological constant,''
  Class.\ Quant.\ Grav.\  {\bf 12}, 2961 (1995)
  [gr-qc/9506019].  
  
  \bibitem{bbo}
  M.~Banados, T.~Brotz and M.~E.~Ortiz,
  ``Boundary dynamics and the statistical mechanics of the (2+1)-dimensional black hole,''
  Nucl.\ Phys.\ B {\bf 545}, 340 (1999)
  [hep-th/9802076].  
  
  \bibitem{hms}
  M.~Henneaux, L.~Maoz and A.~Schwimmer,
  ``Asymptotic dynamics and asymptotic symmetries of three-dimensional extended AdS supergravity,''
  Annals Phys.\  {\bf 282}, 31 (2000)
  [hep-th/9910013].  
  
  \bibitem{car}
S. Carlip, ``Conformal Field Theory, ``$(2+1)$-Dimensional Gravity, and the BTZ Black Hole,'' gr-qc/0503022.  

\bibitem{mart} 
  E.~J.~Martinec,
  ``Conformal field theory, geometry, and entropy,''
  hep-th/9809021.  
  
  \bibitem{yin}
  X.~Yin,
  ``On Non-handlebody Instantons in 3D Gravity,''
  JHEP {\bf 0809}, 120 (2008)
  [arXiv:0711.2803 [hep-th]].  
  
  \bibitem{k}
  K.~Krasnov and J.~M.~Schlenker,
  ``On the renormalized volume of hyperbolic 3-manifolds,''
  Commun.\ Math.\ Phys.\  {\bf 279}, 637 (2008)
  [math/0607081 [math-dg]];
  ``The Weil-Petersson metric and the renormalized volume of hyperbolic 3-manifolds,''
  arXiv:0907.2590 [math.DG];
  K.~Krasnov,
  ``Holography and Riemann surfaces,''
  Adv.\ Theor.\ Math.\ Phys.\  {\bf 4}, 929 (2000)
  [hep-th/0005106].  
 
  \bibitem{wol-mirz} 
  S.~A.~Wolpert,
 ``Lectures and notes: Mirzakhani's volume recursion and approach for the Witten-Kontsevich theorem on moduli 
 tautological intersection numbers,'' arXiv:1108.0174 [math.DG].
 
 \bibitem{men}
 P.~Menotti,
  ``Riemann-Hilbert treatment of Liouville theory on the torus,''
  J.\ Phys.\ A {\bf 44}, 115403 (2011)
  [arXiv:1010.4946 [hep-th]]; ``Riemann-Hilbert treatment of Liouville theory on the torus: The general case,''
  J.\ Phys.\ A {\bf 44}, 335401 (2011)
  [arXiv:1104.3210 [hep-th]];
  ``Accessory parameters for Liouville theory on the torus,''
  JHEP {\bf 1212}, 001 (2012)
  [arXiv:1207.6884 [hep-th]];
  ``Hyperbolic deformation of the strip-equation and the accessory parameters for the torus,''
  JHEP {\bf 1309}, 132 (2013)
  [arXiv:1307.0306 [hep-th]].
  
  \bibitem{wolp}
  S. Wolpert, ``The Weil-Petersson metric geometry," arXiv:0801.0175 [math.DG].
  
  \bibitem{dey}
  R. Dey, ``Hyperk\"ahler quantization of the Hitchin system and Chern-Simons theory with complex gauge group," 
  Adv.\ Theor.\ Math.\ Phys.\  {\bf 11}, 819 (2007). 
  
  \bibitem{mirz} 
  M.~Mirzakhani, ``Weil-Petersson volumes and intersection theory on the moduli space of curves,''
  J.\ Am.\ Math.\ Soc.\  {\bf 20}, no. 01, 1 (2007);
  ``Weil-Petersson Volumes and Intersection Theory on the Moduli Space of Curves," J. \ Amer. \ Math. \ Soc., 
  {\bf 20}(2007) 1-23.
    
  \bibitem{dvv}
  R.~Dijkgraaf, H.~L.~Verlinde and E.~P.~Verlinde,
  ``Notes on topological string theory and 2-D quantum gravity,''
  PUPT-1217, IASSNS-HEP-90-80, C90-04-23, C90-05-27.    
  
  \bibitem{wol-WP}
  S.~A.~Wolpert,  ``On the Weil-Petersson geometry of the moduli space of curves." Am.\ J.\ Math., 969 (1985).  
  
  \bibitem{mazz}
  R.~Mazzeo and J.~Svoboda, ``Asymptotics of the Weil-Petersson metric," arXiv:1503.02365 [math.DG].  
  
  \bibitem{wol-lap} S.~A.~Wolpert, ``Asymptotics of the Spectrum and the Selberg Zeta Function on the Space
of Riemann surfaces,� Commun. \ Math. \ Phys. \ {\bf 112}, 283 (1987)  
  
  \bibitem{cmnp} 
  A.~G.~Cohen, G.~W.~Moore, P.~C.~Nelson and J.~Polchinski,
  ``An Off-Shell Propagator for String Theory,''
  Nucl.\ Phys.\ B {\bf 267}, 143 (1986).  
  
  \bibitem{iengo}
  E.~Gava and R.~Jengo,
  ``On The Cosmological Constant In The Heterotic String Theory,''
  Phys.\ Lett.\ B {\bf 187}, 22 (1987).    
  
  \bibitem{dk} 
  J.~Distler and H.~Kawai,
  ``Conformal Field Theory and 2D Quantum Gravity Or Who's Afraid of Joseph Liouville?,''
  Nucl.\ Phys.\ B {\bf 321}, 509 (1989).  
  
  \bibitem{had}
  L.~Hadasz, Z.~Jaskolski and P.~Suchanek,
  ``Recursive representation of the torus 1-point conformal block,''
  JHEP {\bf 1001}, 063 (2010)
  [arXiv:0911.2353 [hep-th]].
  
   \bibitem{w-ana}
  E.~Witten,
  ``Analytic Continuation Of Chern-Simons Theory,''
  arXiv:1001.2933 [hep-th].
   
 \bibitem{ah}
  K.~Amano and S.~Higuchi,
  ``Topology change in ISO(2,1) Chern-Simons gravity,''
  Nucl.\ Phys.\ B {\bf 377}, 218 (1992)
  [hep-th/9201075].  
  
  \bibitem{gawi}
  D.~Gaiotto and E.~Witten,
  ``Knot Invariants from Four-Dimensional Gauge Theory,''
  Adv.\ Theor.\ Math.\ Phys.\  {\bf 16}, no. 3, 935 (2012)
  [arXiv:1106.4789 [hep-th]].  
  
  \bibitem{caco}
  S.~Carlip and R.~Cosgrove,
  ``Topology change in (2+1)-dimensional gravity,''
  J.\ Math.\ Phys.\  {\bf 35}, 5477 (1994)
  [gr-qc/9406006].  
  
  \bibitem{strom} 
  A.~Strominger,
  ``Black hole entropy from near horizon microstates,''
  JHEP {\bf 9802}, 009 (1998)
  [hep-th/9712251].  
  
  \bibitem{cardy} 
  J.~L.~Cardy,
  ``Operator Content of Two-Dimensional Conformally Invariant Theories,''
  Nucl.\ Phys.\ B {\bf 270}, 186 (1986).  
  
  \bibitem{m-w}
  A.~Maloney and E.~Witten,
  ``Quantum Gravity Partition Functions in Three Dimensions,''
  JHEP {\bf 1002} (2010) 029
  [arXiv:0712.0155 [hep-th]].  
  
  \bibitem{kp1}
  J.~Kim and M.~Porrati,
  ``Long string dynamics in pure gravity on AdS$_{3}$,''
  J.\ Exp.\ Theor.\ Phys.\  {\bf 120}, no. 3, 477 (2015)
  [arXiv:1410.3424 [hep-th]].  
  
  \bibitem{kp2}
  J.~Kim and M.~Porrati,
  ``More on long string dynamics in gravity on AdS$_3$ : Spinning strings and rotating BTZ black holes,''
  Phys.\ Rev.\ D {\bf 91}, no. 12, 124061 (2015)
  [arXiv:1503.06875 [hep-th]].  
  
  \bibitem{bt}
  J.~D.~Brown and C.~Teitelboim,
  ``Dynamical Neutralization of the Cosmological Constant,''
  Phys.\ Lett.\ B {\bf 195}, 177 (1987);
  ``Neutralization of the Cosmological Constant by Membrane Creation,''
  Nucl.\ Phys.\ B {\bf 297}, 787 (1988).  
  
  \bibitem{fqs}
  D.~Friedan, Z.~A.~Qiu and S.~H.~Shenker,
  ``Conformal Invariance, Unitarity and Two-Dimensional Critical Exponents,''
  Phys.\ Rev.\ Lett.\  {\bf 52}, 1575 (1984).
  
  \bibitem{CFPT}
  A.~Campoleoni, S.~Fredenhagen, S.~Pfenninger and S.~Theisen,
  ``Asymptotic symmetries of three-dimensional gravity coupled to higher-spin fields,''
  JHEP {\bf 1011}, 007 (2010)
  [arXiv:1008.4744 [hep-th]].  
  
  \bibitem{gg}
  M.~R.~Gaberdiel and R.~Gopakumar,
  ``An $AdS_3$ Dual for Minimal Model CFTs,''
  Phys.\ Rev.\ D {\bf 83}, 066007 (2011)
  [arXiv:1011.2986 [hep-th]];
  `Minimal Model Holography,''
  J.\ Phys.\ A {\bf 46}, 214002 (2013)
  [arXiv:1207.6697 [hep-th]].  
 
 \bibitem{terayama} 
  Y.~Terashima and M.~Yamazaki,
  ``SL(2,R) Chern-Simons, Liouville, and Gauge Theory on Duality Walls,''
  JHEP {\bf 1108}, 135 (2011)
    
  \end{thebibliography}
 \end{document}